\documentclass[sigconf]{acmart}

\usepackage{enumitem}
\usepackage{algorithm}
\usepackage[noend]{algpseudocode}
\usepackage{multirow}
\usepackage{graphicx}
\usepackage{subcaption}
\usepackage{caption}

\catcode`|=\active
\def|#1|{{\texttt{#1}}}

\newcommand{\multirowcell}[2]{%
    \begin{tabular}{@{}#1@{}} #2 \end{tabular}%
}

\AtBeginDocument{%
  }

\copyrightyear{2025}
\acmYear{2025}
\setcopyright{cc}
\setcctype{by}
\acmConference[MICRO '25]{58th IEEE/ACM International Symposium on
Microarchitecture}{October 18--22, 2025}{Seoul, Republic of Korea}
\acmBooktitle{58th IEEE/ACM International Symposium on Microarchitecture
(MICRO '25), October 18--22, 2025, Seoul, Republic of
Korea}\acmDOI{10.1145/3725843.3762817}
\acmISBN{979-8-4007-1573-0/2025/10}

\begin{document}

\title{StreamTensor: Make Tensors Stream in Dataflow Accelerators for LLMs}
\author{Hanchen Ye}
\authornote{Work was done during an internship at Inspirit IoT, Inc.}
\affiliation{%
    \institution{University of Illinois Urbana-Champaign}
    \city{Urbana}
    \state{Illinois}
    \country{USA}
}
\email{hanchen8@illinois.edu}

\author{Deming Chen}
\affiliation{%
    \institution{Inspirit IoT, Inc.}
    \city{Champaign}
    \state{Illinois}
    \country{USA}
}
\affiliation{%
    \institution{University of Illinois Urbana-Champaign}
    \city{Urbana}
    \state{Illinois}
    \country{USA}
}
\email{deming.chen@inspirit-iot.com}




\begin{abstract}

Efficient execution of deep learning workloads on dataflow architectures is crucial for overcoming memory bottlenecks and maximizing performance. While streaming intermediate results between computation kernels can significantly improve efficiency, existing approaches struggle with inter-kernel correlations, external memory access management, and buffer optimization. In this work, we propose StreamTensor, a compiler framework that automatically constructs and optimizes stream-based dataflow accelerators. StreamTensor introduces a novel iterative tensor type system to explicitly encode stream layouts, enabling seamless kernel fusion, buffer allocation, and memory optimization. By systematically exploring three hierarchical design spaces, including tensor tiling, kernel fusion, and resource allocation, StreamTensor balances computational intensity, memory efficiency, and data streaming to maximize performance. Based on FPGA evaluations on Large Language Models~(LLM), StreamTensor achieves up to 0.76x and 0.64x lower latency compared to the state-of-the-art FPGA LLM accelerators and GPUs, and up to 1.99x higher energy efficiency compared to GPUs, making it a promising approach for scalable dataflow-based deep learning acceleration.

\end{abstract}

\maketitle

\section{Introduction}

\subsection{Dataflow Architecture}

Dataflow architecture, as an alternative to Von Neumann-style architectures such as the NVIDIA H100~\cite{choquette2023NVIDIA} and Google TPUv4~\cite{jouppi2023tpu}, is increasingly adopted and studied to overcome the memory wall in emerging AI applications, such as Large Language Models (LLM). Because of LLMs' autoregressive nature, the decoding stage is highly memory-bound, demanding more memory-efficient architectures. AMD Versal~\cite{gaide2019xilinx}, Sambanova SN40L~\cite{prabhakar2024sambanova}, and IBM AIU~\cite{burns2022meet} are commercial AI accelerators with reconfigurable dataflow architectures; many studies~\cite{prabhakar2017plasticine,nowatzki2017stream,chen2024understanding} have also demonstrated the latency and energy efficiency advantages of dataflow architecture.

\begin{figure}[t]
    \centering
    \includegraphics[width=\linewidth]{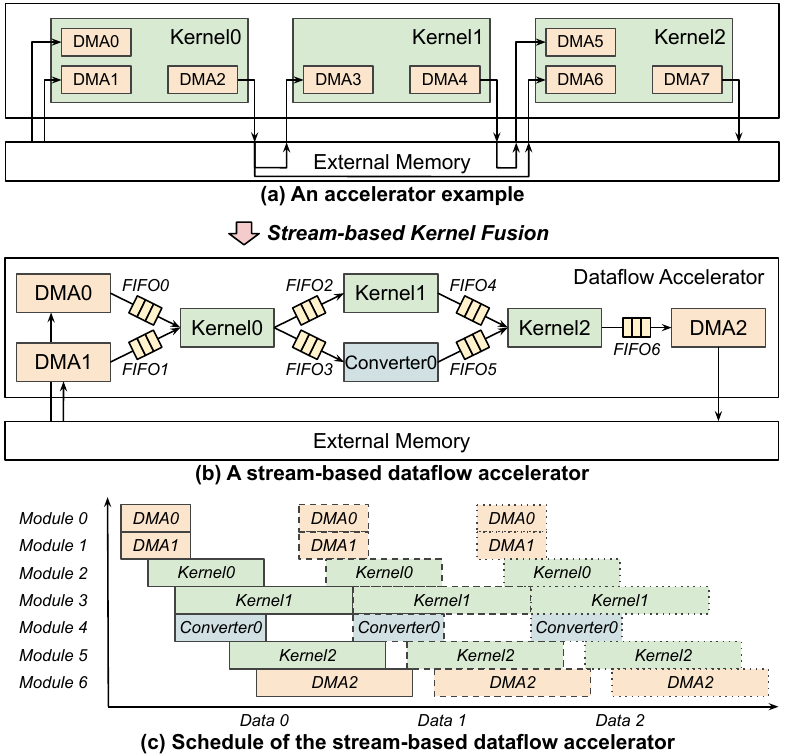}
    \caption{Computation pattern of dataflow accelerators.}
    \label{fig:dataflow}
\end{figure}

Figure~\ref{fig:dataflow} shows the typical computation pattern of dataflow accelerators. As shown in Figure~\ref{fig:dataflow}(b), a dataflow accelerator contains the following on-chip components:

\begin{enumerate}[leftmargin=*]
    \item \textbf{Kernel}: Computes an operator or coarse-grained \emph{task}~(e.g., matrix multiply) using a parallel processor~(e.g., a systolic array), and provides stream interfaces for input and output.
    \item \textbf{Token}: Atomic element communicated between kernels.
    \item \textbf{First-in First-out~(FIFO)}: Holds accumulated stream tokens to balance different token rates of the producer and consumer, and avoids deadlock or unnecessary kernel stalls.
    \item \textbf{Stream Layout Converter}: Converts stream layout on-the-fly to accommodate different computation patterns of producer and consumer kernels through a local ping-pong buffer.
    \item \textbf{Direct Memory Access~(DMA)}: Communicates with external memory, and converts memory-mapped interfaces to stream interfaces or vice versa.
\end{enumerate}

Kernels may be designed using \emph{dataflow} circuits through dynamic scheduling~\cite{josipovic2018dynamically}, or may adopt different \emph{dataflow} strategies (e.g., input stationary) for efficient on-chip data reuse~\cite{chen2016eyeriss}. Although using the same terminology, these \emph{dataflow} concepts are conceptually orthogonal to the dataflow architecture and accelerators discussed in this paper.

The key idea of dataflow architecture is to stream intermediate results between kernels through on-chip FIFOs instead of triggering frequent external memory accesses. For example, in Figure~\ref{fig:dataflow}(b), the intermediate results produced by \emph{Kernel0} are streamed directly to \emph{Kernel1} and \emph{Converter0} without going through external memory, as in Figure~\ref{fig:dataflow}(a). Following the convention proposed in~\cite{prabhakar2024sambanova}, we refer to enabling streaming between dataflow kernels as \emph{stream-based kernel fusion}. Additionally, as illustrated in Figure~\ref{fig:dataflow}(c), the schedule of the dataflow accelerator allows \emph{Kernel1} and \emph{Converter0} to start execution before \emph{Kernel0} completes. This overlapped execution can significantly improve both the overall throughput and latency.

\begin{figure}[t]
    \centering
    \includegraphics[width=\linewidth]{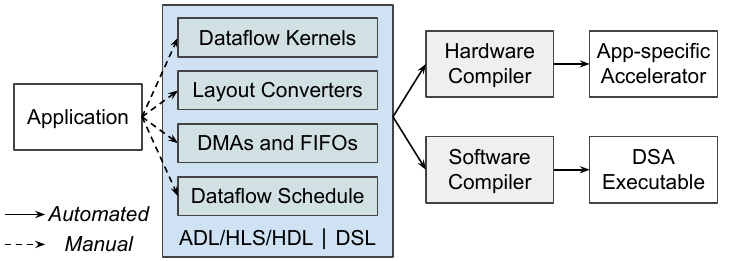}
    \caption{Current paradigm of dataflow accelerator design.}
    \label{fig:current_paradigm}
\end{figure}

\subsection{Dataflow Accelerator Programming}
\label{sec:programming}

Figure~\ref{fig:current_paradigm} shows the current paradigm of dataflow accelerator programming. As dataflow accelerators generally fall into two categories, application-specific accelerators and domain-specific accelerators~(DSAs), we discuss each separately.

\subsubsection{Application-specific Accelerator} 

In this category, the dataflow components and schedule are tailored for a single application. Thus, \emph{programming} typically refers to the \emph{design} or \emph{generation} of architecture and microarchitecture. Traditionally, Hardware Description Languages~(HDLs), High-level Synthesis~(HLS), and meta-HDLs like Chisel~\cite{bachrach2012chisel} are used for this purpose~\cite{chen2005xpilot, zhang2018dnnbuilder,chi2018soda,sarkar2023flowgnn}. More recently, Accelerator Design Languages~(ADLs) have emerged to improve productivity~\cite{chen2024allo,durst2020type,thomas2020fleet}, introducing typing systems and primitives to describe computation, memory layout, and dataflow schedules. As shown in Figure~\ref{fig:current_paradigm}, existing solutions require manual effort to convert applications into dataflow schedules and components, which are then passed to HLS, meta-HDL transpilers, or vendor EDA tools for hardware generation. While ADLs and HLS frameworks incorporate Design Space Exploration~(DSE)~\cite{koeplinger2016automatic,koeplinger2018spatial,ben2019stateful,ye2022scalehls,agostini2022mlir,zhang2024optimizing}, these efforts focus mainly on optimizing individual kernels.

\subsubsection{Dataflow DSA}

DSAs are designed to efficiently perform computations for a particular class of applications or a specific domain, rather than being a general-purpose processor. DSAs are often realized using Coarse-grained Reconfigurable Architecture~(CGRA)-like architectures~\cite{gaide2019xilinx,prabhakar2024sambanova,prabhakar2017plasticine,nowatzki2017stream}, where on-chip resources are reconfigured to implement different dataflow designs. Modern DSAs are programmed using C/C++ primitives~\cite{gaide2019xilinx,zhuang2023charm,zhuang2024charm} or Domain-specific Languages (DSLs), such as Spatial~\cite{koeplinger2018spatial}, Halide~\cite{ragan2013halide}, and TVM~\cite{chen2018tvm}, to generate domain-optimized code. As illustrated in Figure~\ref{fig:current_paradigm}, developers must manually transform applications into logical components using these DSLs or APIs. Software compilers then map them to physical resources and generate the final binaries for on-chip execution. While these DSLs often provide auto-tuning capabilities for dataflow kernels, their primary focus is on optimizing individual kernels instead of the entire dataflow application, leaving substantial performance gains unrealized.

\subsection{Pitfalls}
\label{sec:pitfalls}


\subsubsection{Pitfall 1: Inter-kernel Correlation}

Prior works~\cite{ye2022scalehls,ye2024hida} show that inter-kernel correlation can affect accelerator performance. Since kernels execute in a pipelined manner, their latencies must be balanced for optimal throughput. Moreover, buffer-connected kernels need aligned parallelization strategies to avoid inefficient memory use. However, previous work only considered ping-pong buffers, which support memory-mapped access. FIFOs are more restrictive, as data must be pushed/pulled in order. This introduces the following challenges for each kernel:
\begin{enumerate}[leftmargin=*]
    \item \emph{Tiling}: Choosing tile sizes that enable streaming, minimize local buffering, and preserve memory efficiency.
    \item \emph{Permutation}: Reordering loops to reduce memory utilization during data streaming.
    \item \emph{Vectorization}: Selecting unrolling strategies to balance latency and improve streaming efficiency.
\end{enumerate}

These decisions are interdependent across kernels, making global optimization challenging for analytical models or manual design.

\subsubsection{Pitfall 2: External Memory Access}

Most existing compilers~\cite{ye2022scalehls,agostini2022mlir,ye2024hida,zhao2022polsca,zhang2024optimizing,basalama2025stream} assume that all data fits on-chip, which is unrealistic for large applications. When off-chip memory is involved, each DMA must address the following issues:
\begin{enumerate}[leftmargin=*]
    \item How to overlap memory access with kernel execution?
    \item What data layout best matches the streaming pattern?
    \item How to pack/vectorize data to maximize bandwidth?
\end{enumerate}

These require nontrivial pattern analysis and are error-prone when handled manually. DMA design is also tightly coupled with kernel tiling and scheduling, compounding the complexity.

\subsubsection{Pitfall 3: Stream-based Kernel Fusion}

The goal of stream-based kernel fusion is to stream all intermediate results on-chip, limiting external memory use to inputs and outputs. However, producer and consumer kernels often have incompatible stream layouts due to different computation patterns. This requires:
\begin{enumerate}[leftmargin=*]
    \item Checking layout compatibility between kernels.
    \item Generating minimal on-the-fly stream layout converters.
    \item Ensuring the converter fits within available on-chip memory.
\end{enumerate}

These steps involve complex pattern analysis and require a global view of the system, making manual solutions impractical.

\subsubsection{Pitfall 4: FIFO Sizing}

As shown in Figure~\ref{fig:dataflow}, if \emph{Kernel1} is slower than \emph{Converter0}, FIFOs may overflow or underflow, leading to a stall cascade and eventual deadlock. Though dynamic scheduling solutions exist~\cite{josipovic2021buffer}, coarse-grained accelerators still rely on manual sizing~\cite{chen2024allo,chen2024understanding}, which does not scale to a large number of FIFOs. A recent automated approach~\cite{honorat2024automated} uses simulation to determine FIFO sizes, but it is time-consuming and lacks scalability.

\begin{figure}[t]
    \centering
    \includegraphics[width=\linewidth]{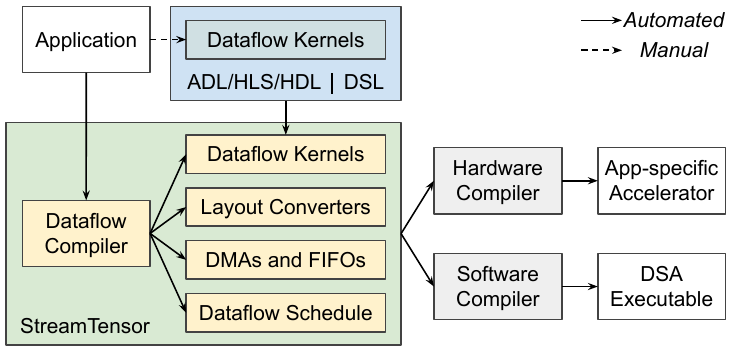}
    \caption{Proposed paradigm of dataflow accelerator design.}
    \label{fig:proposed_paradigm}
\end{figure}

\subsection{Our Proposal}
\label{sec:proposal}

Due to the pitfalls described in Section~\ref{sec:pitfalls}, the current paradigm shown in Figure~\ref{fig:current_paradigm} is difficult to scale up to large dataflow accelerators. Therefore, we propose a shift in the design paradigm shown in Figure~\ref{fig:proposed_paradigm}. We do not advocate for full automation, as ADL/HLS/HDL or DSLs remain essential for designing individual dataflow kernels, such as local buffers and vectorization. However, once individual kernels are designed or generated, we argue that compilers should automatically generate the dataflow schedule, assemble the kernels into an application-level dataflow accelerator, and resolve the pitfalls identified in Section~\ref{sec:pitfalls} algorithmically. This is analogous to the GPU software ecosystem, where DSLs like CUDA and Triton~\cite{tillet2019triton} are used to design or auto-tune individual GPU kernels, while kernel assembly and scheduling are handled automatically by compilers, resulting in a programming paradigm that is both efficient and scalable.

In this spirit, we propose \emph{StreamTensor}, a compiler that enables automatic tensor streaming in dataflow architectures. This paper describes how each pitfall is addressed in a systematic and hierarchical manner. As a pioneering work, StreamTensor proposes algorithmic solutions for each challenge and demonstrates their effectiveness through large benchmarks. While these solutions may not be optimal, they clearly expose well-defined optimization subproblems and enable co-optimization opportunities across different design spaces.
Overall, this paper makes the following contributions:
\begin{enumerate}[leftmargin=*]
    \item We propose StreamTensor, the first PyTorch-to-device dataflow compiler that automatically generates stream-based dataflow accelerators and their corresponding runtime systems.
    \item We propose an iterative tensor~(|itensor|) type that systematically encodes the stream information for the first time. This typing system forms the foundation for stream-based kernel fusion and dataflow component generation, improving the scalability and productivity of dataflow accelerator design.
    \item We propose three design spaces, including tensor tiling space, kernel fusion space, and resource allocation space, that cover the sophisticated design space of dataflow architecture in an algorithmic and hierarchical manner. We further propose an exploration algorithm for each design space
    to reduce resource utilization and improve latency and throughput.
    \item We propose a piecewise function-based token behavior model that transforms the dataflow FIFO sizing problem of dataflow accelerators into a scheduling problem. We further propose a linear programming~(LP) algorithm to solve this problem, reducing resource utilization while avoiding deadlock.
    \item We evaluate StreamTensor on FPGA platforms with LLMs and observe up to 0.76x and 0.64x lower latency compared to the state-of-the-art FPGA LLM accelerators and GPUs, and up to 1.99x higher energy efficiency compared to GPUs.
\end{enumerate}

\begin{figure*}
    \centering
    \includegraphics[width=0.96\textwidth]{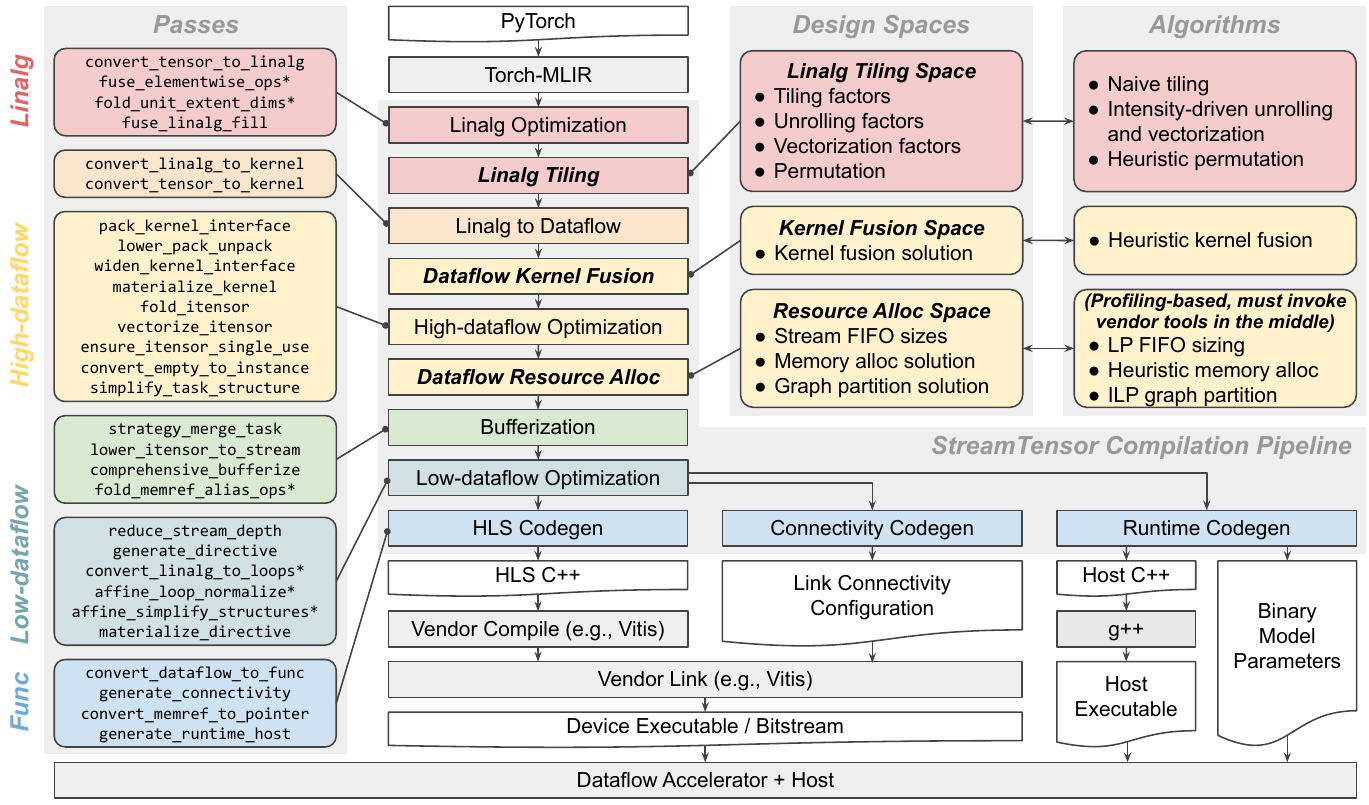}
    \caption{Proposed StreamTensor framework.}
    \label{fig:framework}
\end{figure*}

\section{StreamTensor Framework}

StreamTensor is a compilation framework designed to transform PyTorch models into optimized dataflow implementations. It is built upon the MLIR~\cite{lattner2021mlir} compilation framework. The overall architecture of StreamTensor is depicted in Figure~\ref{fig:framework}. The compilation process begins with a PyTorch model from Torch-MLIR~\cite{torch_mlir_github} and proceeds through several stages. Initially, tensor operations are converted into a structured Intermediate Representation (IR) using MLIR's built-in Linear Algebra (Linalg) operations. This IR is then optimized by MLIR's Linalg passes like element-wise operation fusion. StreamTensor subsequently applies Design Space Exploration (DSE) algorithms to determine optimal tiling strategies, considering factors such as tile sizes, unrolling factors, and permutations based on computational patterns. The Linalg IR is then transformed into a dataflow IR, where computations are organized as hierarchical tasks. All dataflow components, including DMAs, stream layout converters, and FIFOs, are generated during this stage. Critical optimizations are also performed here, such as stream-based kernel fusion to minimize external memory access and FIFO sizing to balance producer-consumer executions. In the final stages, StreamTensor generates hardware-specific code and a host runtime. StreamTensor handles memory allocation, stream connectivity, and directive materialization, which allows vendor compilers like HLS to generate the target dataflow architectures. Concurrently, it produces host runtime code that manages data transfer, kernel execution, and synchronization between the host CPU and the dataflow accelerator.


\begin{figure*}
    \centering
    \includegraphics[width=\linewidth]{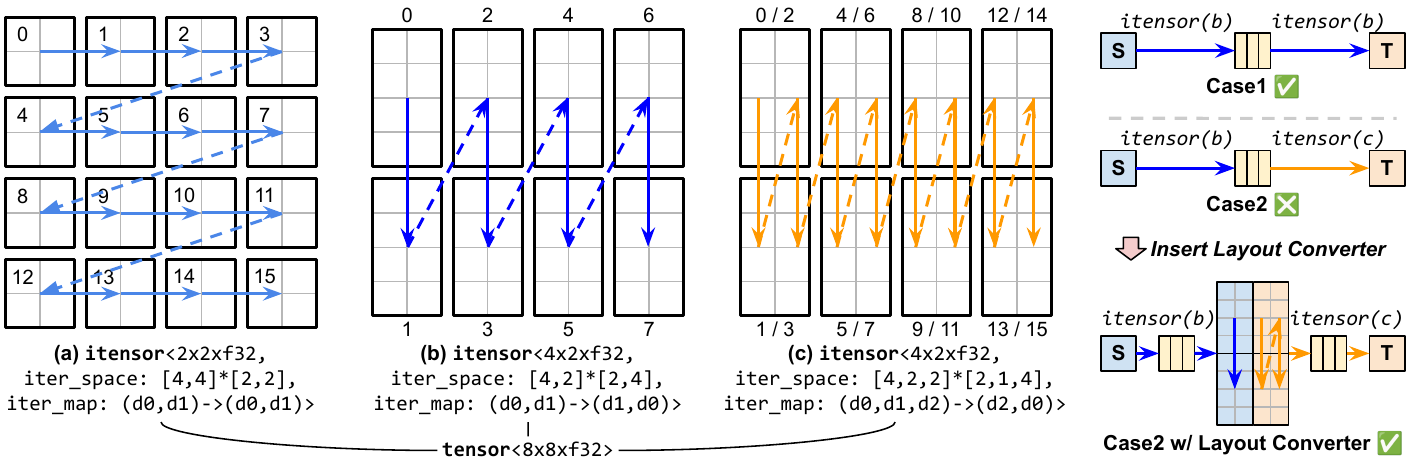}
    \caption{Iterative tensor (\texttt{|itensor|}) typing system.}
    \label{fig:itensor}
\end{figure*}

\section{Intermediate Representation}

\subsection{Typing System}
\label{sec:types}

StreamTensor introduces a typing system to enable efficient verification and optimization of the IR. Through StreamTensor's dedicated type and operation verifiers, the typing system helps ensure the IR's validity after any transformation pass is applied.

\subsubsection{Motivation}

Traditionally, |tensor| type encodes a data type and a list of integers representing its shape~\cite{chen2018tvm,ragan2013halide,lattner2021mlir,hagedorn2023graphene}. Tensors can be accessed in a memory-mapped manner, e.g., a slice can be extracted or inserted based on offsets and its shape.
However, dataflow kernels communicate via FIFOs, which enforce a strict access order and follow a streamed access pattern rather than a memory-mapped one. Consequently, traditional tensor types may fail to ensure correctness in dataflow communication. Even when a producer and a consumer share the same tensor type, the stream access order may remain ambiguous, causing unintended behaviors. For example, in Graphene~\cite{hagedorn2023graphene}, the tensor type only encodes memory-mapped layout. As a result, a mismatch between a producer's row-major stream generation and a consumer's column-major expectation, when both operate on the same tensor type, leads to incorrect data interpretation and logical corruption. Therefore, although existing solutions are sufficient for Linalg-level optimizations like tiling, they are error-prone and unscalable for generating dataflow components and applying dataflow optimizations.

\begin{table*}
    \centering
    \small
    \caption{Iterative tensor (|itensor|) operations.}
    \label{tab:itensor_ops}
    \begin{tabular}{cccl}
        \toprule
        \textbf{Operation} & \textbf{Operands} & \textbf{Results} & \textbf{Description} \\
        \midrule
        |itensor\_empty| & - & |result| (|itensor|) & A placeholder representing an empty |itensor|. \\
        \hline
        |itensor\_instance| & - & |result| (|itensor|) & An instance of |itensor| that will be lowered to a FIFO. \\
        \hline
        |itensor\_read| & \multirowcell{c}{|source| (|itensor|) \\ |init| (|tensor|)} & |value| (|any|) & \multirowcell{l}{Read (pull) |value| from |itensor| |source|. Operand |init| is the destination to \\ store |value| when |value| is |tensor| type.} \\
        \hline
        |itensor\_write| & \multirowcell{c}{|value| (|any|) \\ |dest| (|itensor|)} & |result| (|itensor|) & \multirowcell{l}{Write (push) |value| into |itensor| |dest| (destination). |itensor| |result| is the \\ written/pushed |itensor| |dest|.} \\
        \hline
        |itensor\_cast| & |source| (|itensor|) & |result| (|itensor|) & Cast from |source| to |result| without changing stream layout. \\
        \hline
        \multirowcell{c}{|itensor\_| \\ |reassociate|} & |source| (|itensor|) & |result| (|itensor|) & \multirowcell{l}{Reassociate the element shape and/or iteration space of |source| into |result|. \\ Typically lowered from |tensor| |expand\_shape| or |collapse\_shape|.} \\
        \hline
        \multirowcell{c}{|itensor\_| \\ |converter|} & |source| (|itensor|) & |result| (|itensor|) & \multirowcell{l}{Convert stream layout from |source| to |result| with a local ping-pong buffer. \\ Typically generated during dataflow kernel fusion.} \\
        \hline
        |itensor\_chunk| & |source| (|itensor|) & |results| ([|itensor|]) & \multirowcell{l}{Chunk |source| into variadic |results|. Lowered from |tensor| |chunk|.} \\
        \hline
        |itensor\_concat| & |sources| ([|itensor|]) & |result| (|itensor|) & \multirowcell{l}{Concat variadic |sources| into |result|. Lowered from |tensor| |concat|.} \\
        \hline
        |itensor\_fork| & |source| (|itensor|) & |results| ([|itensor|]) & Fork |source| into variadic number of duplicated |results|. \\
        \hline
        |itensor\_join| & |sources| ([|itensor|]) & |result| (|itensor|) & Join variadic number of |sources| into |result| through round-robin. \\
        \bottomrule
    \end{tabular}
\end{table*}

\subsubsection{Iterative Tensor Type}
\label{sec:itensor_type}

To address this, we propose a new |itensor| type that explicitly encodes stream layout information, making type-based verification and optimization both possible and efficient. Figure~\ref{fig:itensor} shows three examples of |itensor|s converted from the same |tensor| with type |tensor<8x8xf32>|. To convert a tensor to an |itensor|, we first partition it into identical tensor slices or vectors. For example, in Figure~\ref{fig:itensor}(b), the tensor is partitioned into eight tensor slices of shape |4x2|. These slices are then accessed iteratively within a defined iteration space, typically nested loops. The iteration space is defined by two lists: tripcounts and step sizes. In Figure~\ref{fig:itensor}(b), the iteration space is |[4,2]*[2,4]|, which produces iteration indices |[0,0]|, |[0,4]|, |[2,0]|, |[2,4]|, etc. The mapping from iteration space to data space is specified by an affine map—for example, |(d0,d1)->(d1,d0)| in Figure~\ref{fig:itensor}(b), which transposes the iteration indices. Thus, the data access indices become |[0,0]|, |[4,0]|, |[0,2]|, |[4,2]|, etc., reflecting this transposition as shown in Figure~\ref{fig:itensor}(b). In |itensor|, tensor slices can be accessed multiple times, with the pattern explicitly encoded in the iteration map. For instance, in Figure~\ref{fig:itensor}(c), the iteration space is |[4,2,2]*[2,1,4]| and the iteration map is |(d0,d1,d2)->(d2,d0)|, where dimension |d1| does not correspond to any data dimension. As |d1| iterates from 0 to 1, all less significant dimensions (like |d2|) are reiterated. Consequently, the corresponding data dimensions (e.g., row dimension) are also re-accessed, producing indices like |[0,0]|, |[4,0]|, |[0,0]|, |[4,0]|, |[0,2]|, etc., for tensor slices of shape |4x2|. By encoding the element shape, iteration space, and iteration map in the |itensor| type, the stream pattern of a dataflow kernel can be uniquely determined. When the |itensor| types of a producer and consumer match, streaming communication can be safely established between them (\emph{Case1} of Figure~\ref{fig:itensor}). Otherwise, a stream layout converter must be inserted in between (\emph{Case2} of Figure~\ref{fig:itensor}), and the minimal ping-pong buffer size for layout conversion can be analytically inferred from the |itensor| types. The details of layout converter generation will be discussed in Section~\ref{sec:converter_generation}. Due to the lack of stream information, existing tensor-based typing systems are not sufficient for stream-based kernel fusion, limiting their usability in stream-based dataflow optimizations.

\subsubsection{Stream Type}
\label{sec:stream_type}

In traditional tensor compilers, high-level tensor IR must be \emph{bufferized} into a low-level memory/buffer IR to enable low-level optimizations and code generation. Following this convention, we propose a |stream| type, which is lowered from |itensor| type during bufferization. Unlike immutable |itensor| objects, |stream| objects represent hardware FIFOs and support mutation through operations such as stream reads and writes. The |stream| type encodes only the data type and FIFO depth, while the stream layout information is stripped during bufferization. As a result, dataflow component generation and optimization must be completed at the |itensor| level IR. After bufferization, the |stream| IR is reserved for lower-level hardware/runtime optimizations and code generation.

\begin{table*}[t]
    \centering
    \small
    \caption{Stream (|stream|) and buffer operations.}
    \label{tab:stream_ops}
    \begin{tabular}{cccl}
        \toprule
        \textbf{Operation} & \textbf{Operands} & \textbf{Results} & \textbf{Description} \\
        \midrule
        |itensor\_to\_stream| & |source| (|itensor|) & |result| (|stream|) & Convert |itensor| |source| to |result|. Must be eliminated during bufferization. \\
        \hline
        |stream\_to\_itensor| & |source| (|stream|) & |result| (|itensor|) & Convert |source| to |itensor| |result|. Must be eliminated during bufferization. \\
        \hline
        |stream| & - & |result| (|stream|) & \multirowcell{l}{A FIFO with a specified depth. Typically lowered from |itensor\_instance|.} \\
        \hline
        |stream\_read| & |source| (|stream|) & |value| (|any|) & \multirowcell{l}{Read (pull) |value| from |source| FIFO. Typically lowered from |itensor\_read|.} \\
        \hline
        |stream\_write| & \multirowcell{c}{|value| (|any|) \\ |dest| (|stream|)} & - & \multirowcell{l}{Write (push) |value| into |dest| FIFO. Typically lowered from |itensor\_write|.} \\
        \hline
        |stream\_cast| & |source| (|stream|) & |result| (|stream|) & Cast from |source| to |result| without changing the stream layout. \\
        \hline
        |buffer| & - & |result| (|memref|) & \multirowcell{l}{A ping-pong (double) buffer. Typically lowered from |tensor\_instance|.} \\
        \bottomrule
    \end{tabular}
\end{table*}

\begin{table*}[t]
    \centering
    \small
    \setlength\tabcolsep{3pt}
    \caption{Structure operations.}
    \label{tab:structure_ops}
    \begin{tabular}{ccccl}
        \toprule
        \textbf{Operation} & \textbf{Operands} & \textbf{Results} & \textbf{Region} & \textbf{Description} \\
        \midrule
        |kernel| & |sources| ([|tensor|]) & |results| ([|tensor|]) & Isolated & \multirowcell{l}{Contains a graph of |task|s. |tensor| |sources| and |results| are implicitly converted \\ to/from |itensor| at the boundary, which will be materialized as DMA tasks.} \\
        \hline
        |task| & \multirowcell{c}{|inits| \\ ([|itensor|/|tensor|])} & \multirowcell{c}{|results| \\ ([|itensor|/|tensor|])} & Transparent & \multirowcell{l}{Contains a graph of operations. Can be nested to form a multi-level dataflow. \\ Outputs are written/pushed into |inits|. |results| are the updated |inits|.} \\
        \hline
        |yield| & \multirowcell{c}{|outputs| \\ ([|itensor|/|tensor|])} & - & - & \multirowcell{l}{Terminator of |kernel| or |task|. Yields |outputs| to the outside of the enclosing \\ |kernel| or |task| region.} \\
        \bottomrule
    \end{tabular}
\end{table*}

\subsection{Operations}
\label{sec:ops}

Built upon the typing system in Section~\ref{sec:types}, StreamTensor introduces |itensor| and |stream| operations to represent different dataflow behaviors. Additionally, structure operations are introduced to represent the multi-level hierarchy of a dataflow accelerator, and are shared by both |itensor| and |stream|-level IRs.

\subsubsection{Iterative Tensor Operations}

Table~\ref{tab:itensor_ops} lists the complete set of operations at the |itensor| level. Overall, these operations are self-explanatory; we highlight those whose semantics are less obvious. |itensor\_write| can be conceptually understood as writing or pushing an element into a FIFO. It is a destination-carried operation, where the destination is an |itensor| passed through a |dest| operand. For example, iteratively writing the |itensor| in Figure~\ref{fig:itensor}(b) (referred to as |itensor(b)|) can be expressed as:

{\small
\begin{verbatim}
%empty = itensor_empty() : itensor(b)
%res0 = scf.for 0 to 8 step 2 iter_args={%arg0 = %empty} {
  %res1 = scf.for 0 to 8 step 4 iter_args={%arg1 = %arg0} {
    %value = ... : tensor<4x2xf32> // %value is defined
    %output = itensor_write %value into %arg1 : ...
    scf.yield %output : itensor(b)
  } : itensor(b)
  scf.yield %res1 : itensor(b)
} : itensor(b)
\end{verbatim}
}

Here, |scf| is an MLIR built-in dialect for structured control flow, including |for| loops. |scf.for| is also destination-carried, where |\%empty| is passed as an argument and iteratively pushed through an |itensor\_write|. Eventually, |\%res0| is returned as the final result. In contrast, |itensor\_read| represents pulling an element from a FIFO. For example, reading |itensor(b)| can be expressed as:

{\small
\begin{verbatim}
%source = ... : itensor(b) // %source is defined
scf.for 0 to 8 step 2 {
  scf.for 0 to 8 step 4 {
    %empty = tensor.empty() : tensor<4x2xf32>
    %value = itensor_read %source init %empty : ...
    ... = ... %value ... // %value is used
} }
\end{verbatim}
}

|itensor\_converter| contains a local ping-pong buffer that performs on-the-fly stream layout conversion. For example, in \emph{Case1} of Figure~\ref{fig:itensor}, the source and target share the same |itensor| type and can connect via a FIFO. In \emph{Case2}, they differ, so a converter must be inserted. A minimum |8x2| ping-pong buffer is required to accommodate the stream layouts. While the source writes to the ping buffer, the target reads the pong buffer twice, then they swap.




\subsubsection{Stream Operations}

Table~\ref{tab:stream_ops} lists the operations at the |stream| level. These are mostly self-explanatory; we highlight the key difference from |itensor| operations. As discussed in Section~\ref{sec:stream_type}, |stream| objects are mutable, and destination-carried semantics are no longer used. A FIFO push and pull can be written as:

{\small
\begin{verbatim}
%stream = stream() : stream<f32, depth: 32>
scf.for 0 to 8 step 2 {
  scf.for 0 to 8 step 4 {
    %value = ... : f32 // %value is defined
    stream_write %value into %stream : ...  
} }
scf.for 0 to 8 step 2 {
  scf.for 0 to 8 step 4 {
    %value = stream_read %stream : ...
    ... = ... %value ... // %value is used
} }
\end{verbatim}
}

Note that the same |\%stream| is used throughout without creating new duplicates, unlike the destination-carried style of |itensor|. |stream| IR is more efficient for code generation, but complicates define-use analysis. Hence, |itensor| is preferred for high-level dataflow optimization. The correctness of |stream| operations is guaranteed by construction as they are lowered from |itensor| operations, which are strictly verified by the |itensor| typing system.

\begin{figure*}
    \centering
    \includegraphics[width=0.75\linewidth]{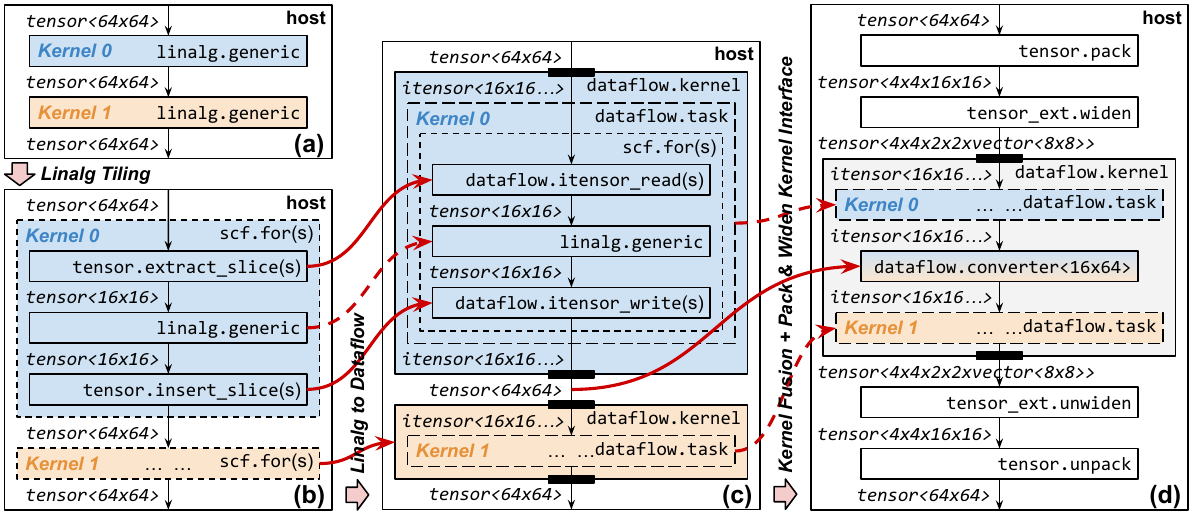}
    \caption{Linalg tiling, Linalg to dataflow conversion, and dataflow kernel fusion. A solid arrow indicates an operation on the left is transformed into the operation on the right, whereas a dashed arrow indicates an operation that remains unchanged.}
    \label{fig:tiling_fusion}
\end{figure*}

\subsubsection{Structure Operations}

While |itensor| and |stream| operations model behavior, structure operations model hierarchy. Table~\ref{tab:structure_ops} lists all the structure operations in StreamTensor. The |kernel| operation represents a dataflow kernel (as in Figure~\ref{fig:dataflow}), containing a graph of |task| operations. It takes |tensor|s as inputs/outputs, which are converted to/from |itensor|s at the boundary. These implicit conversions act as DMAs. Intra-|kernel| uses on-chip streaming, while inter-|kernel| uses external memory. For example:

{\small
\begin{verbatim}
%source = ... : tensor<8x8xf32> // %source is defined
%result = kernel(%arg : itensor<b> = 
                 %source : tensor<8x8xf32>) {
  ... = ... %arg ... // %arg is used
  %output = ... : itensor<c> // %output is defined
  yield %output : itensor<c>
} : tensor<8x8xf32>
\end{verbatim}
}

By converting at the kernel boundary, we avoid explicit DMA handling during kernel fusion, improving transformation efficiency and analyzability. In contrast, the |task| operation is transparent and does not convert types at its boundary. It represents a dataflow task within a kernel and may be nested for hierarchical dataflow designs. At the |itensor| level, |task| is destination-carried where outputs are written into destinations via |inits|, improving the efficiency of define-use analysis. For example:

{\small
\begin{verbatim}
%empty = ... : itensor(b)
%result = task @example inits={%arg = %empty} {
  %value = ... : tensor<4x2xf32> // %value is defined
  %output = itensor_write %value into %arg : ...
  yield %output : itensor(b)
} : itensor(b)
\end{verbatim}
}

After lowering and bufferization, the same code becomes:

{\small
\begin{verbatim}
%stream = stream() : stream<f32, depth: 32>
task @example {
  %value = ... : f32 // %value is defined
  stream_write %value into %stream : ...
}
\end{verbatim}
}

We can observe that |task| combines both |itensor| and |stream| operations, making it a unifying structure abstraction across both IRs that serve different levels of dataflow optimizations. Eventually, all dataflow |task|s are lowered to MLIR built-in |call| and |func| operations for code generation.

\section{Compilation Pipeline}
\label{sec:pipeline}

Building on the type system and operations, we introduce a compilation pipeline that compiles Linalg IR into hardware implementations and a corresponding runtime. All compilation passes are shown in Figure~\ref{fig:framework}.
In this section, we focus on the Linalg-to-dataflow conversion, dataflow kernel fusion, and dataflow optimizations that are unique and essential to understanding the compiler.


\subsection{Linalg to Dataflow}

Figures~\ref{fig:tiling_fusion}(a)-(c) illustrate the Linalg-to-dataflow conversion process. The original Linalg operations (Figure~\ref{fig:tiling_fusion}(a)) are first tiled into Figure~\ref{fig:tiling_fusion}(b), where |scf.for|s represent the loop nests for tiling. In each iteration, |extract\_slice|s extract input tensor tiles to feed into the tiled Linalg operation. After the operation produces output tiles, |insert\_slice|s insert them back into the full tensor. Then, each tiled loop nest is converted into a |kernel| operation in place as shown in Figure~\ref{fig:tiling_fusion}(c). The input and output |tensor|s are converted into/from |itensor|s at the boundary of |kernel|s. The |itensor| types are inferred from:
\begin{enumerate}[leftmargin=*]
    \item The nested |scf.for| loops — iteration tripcounts and step sizes define the |itensor| iteration space.
    \item The |extract\_slice| and |insert\_slice| operations' offsets and sizes — offsets define the iteration mapping, while sizes define the element shape. For example, offsets |[\%iv2, \%iv0]| result in the iteration map |(d0,d1,d2)->(d2,d0)|.
\end{enumerate}

After conversion, |extract\_slice| and |insert\_slice| operations are replaced with |itensor\_read| and |itensor\_write| operations, respectively. The resulting |scf.for| loop nest is wrapped in a |task| to form a single-level dataflow hierarchy: a dataflow kernel containing a dataflow task. By converting the Linalg semantics to dataflow, we open opportunities for subsequent dataflow-oriented transforms and optimizations.


\subsection{Dataflow Kernel Fusion}
\label{sec:fusion}

After all tiled Linalg operations are converted to dataflow kernels, all these kernels initially communicate via traditional |tensor|s, which are eventually stored in external memory. To reduce this communication overhead, StreamTensor applies stream-based kernel fusion. Figures~\ref{fig:tiling_fusion}(c)-(d) show this process. To fuse \emph{Kernel0} and \emph{Kernel1}, we first compare the output |itensor| type of \emph{Kernel0} with the input |itensor| type of \emph{Kernel1}. As described in Section~\ref{sec:itensor_type}, if the types match, we can directly fuse the kernels. If not, we insert a stream layout converter as shown in Figure~\ref{fig:tiling_fusion}(d). The fused kernel comprises two |task|s and a |converter|, all communicating via |itensor|s that will be lowered to on-chip stream FIFOs. The |itensor| typing system enables any dataflow kernels to be fused \emph{by design} at the cost of potential on-chip memory utilization for converters. In Section~\ref{sec:kernel_fusion_space}, we will discuss the exploration of kernel fusion space given memory constraints.

After fusion, StreamTensor applies additional optimization passes to improve the efficiency of external memory access. In particular, |tensor| |pack| and |unpack| operations are inserted before and after the |kernel| to convert between default and tiled memory layouts for burst memory access. For example, with a tiling size of |[16,16]| on a |64x64| tensor, the packed tensor has shape |4x4x16x16|. To maximize the usage of external memory bandwidth, StreamTensor widens the tensor with vectors. For instance, with 512-bit DDR or HBM and |uint8| elements, grouping 64 elements into |vector<64>| fully utilizes the bandwidth. In Figure~\ref{fig:tiling_fusion}, the packed tensor is widened to shape |4x4x2x2xvector<8x8>|. Note that |pack| and |widen| operations are eventually lowered to runtime operations on the host CPU, which prepares data for the accelerator and causes some latency and memory overhead. However, for static tensors (e.g., pre-trained parameters), |pack| and |widen| can fuse directly into these tensors, eliminating any runtime costs. For dynamic tensors (e.g., activations), |pack| and |widen| operations can be folded with their |unpack| and |unwiden| counterparts from the preceding layer via effective Linalg tiling space exploration. As a result, the |pack| and |widen| operations, being necessary only for the model's inputs and outputs, contribute negligible memory and latency overhead at runtime.

\begin{figure*}
    \centering
    \includegraphics[width=\linewidth]{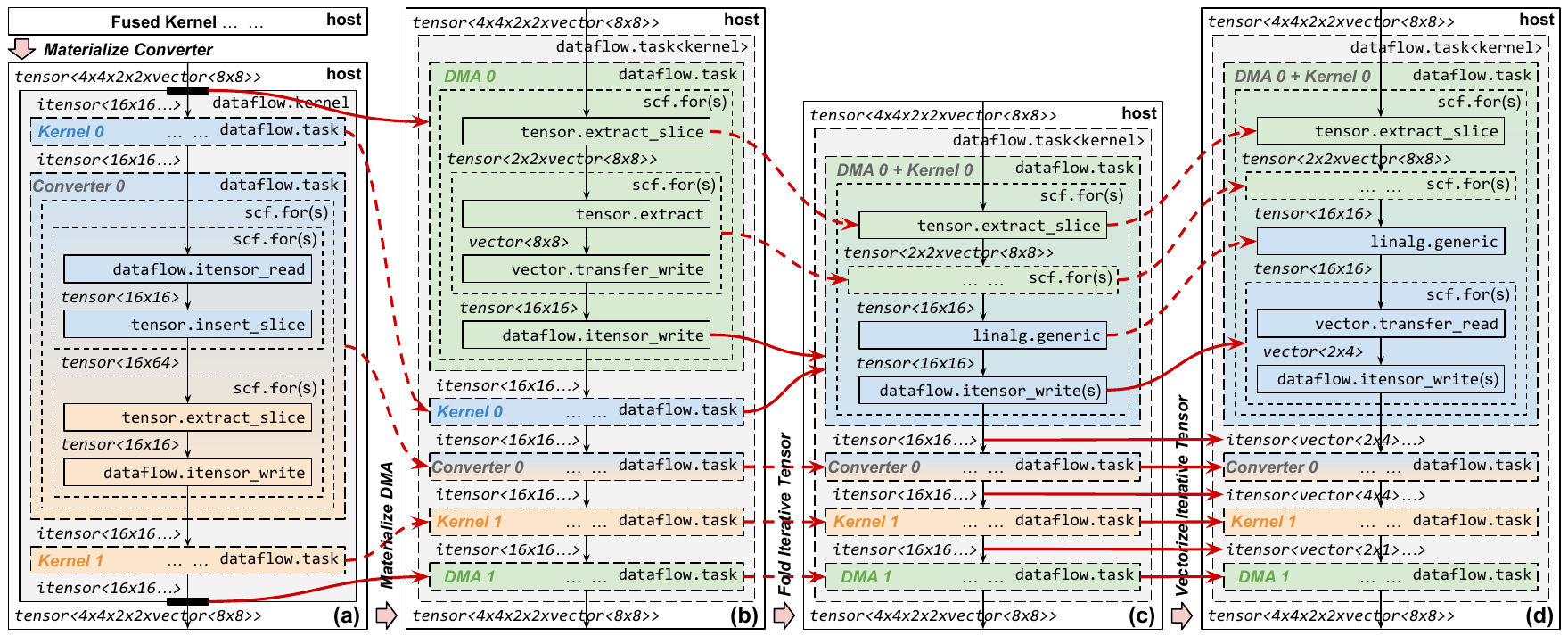}
    \caption{Materialize converter \& DMA, fold |itensor|, and vectorize |itensor|. A solid arrow indicates an operation on the left is transformed into the operation on the right, whereas a dashed arrow indicates an operation that remains unchanged.}
    \label{fig:optimization}
\end{figure*}

\subsection{Dataflow Optimization}

\subsubsection{Materialization}

Figures~\ref{fig:optimization}(a) and (b) illustrate the \emph{materialization} process for converters and DMAs. Materialization involves transforming a high-level dataflow component into its low-level implementation, typically |scf.for| loop nests containing |tensor| and |itensor| operations. Initially, converters are represented by |itensor\_converter|, while DMAs are implicitly handled via |tensor| to or from |itensor| conversions at |kernel| boundaries. This abstraction facilitates kernel fusion and converter optimization. For instance, redundant converters generated for multiple consumers of a producer can be removed using MLIR’s Common Sub-expression Elimination (CSE), which becomes harder after materialization. In contrast, after materialization, all dataflow components are expressed as nested |task|s, making further dataflow optimizations efficient and accessible. For converters, as shown in Figure~\ref{fig:optimization}(a), \emph{Converter0} contains two |scf.for| loop nests connected with a |16x64| ping-pong buffer. These two loop nests are wrapped by a \emph{shared} parent |scf.for| loop to iterate through the original full |64x64| tensor. Therefore, the |16x64| ping-pong buffer is reused four times, effectively reducing on-chip memory resource utilization by a factor of four. In Section~\ref{sec:kernel_fusion_space}, we will discuss how the ping-pong buffer shape and shared loops are inferred from |itensor| types.

For DMAs, as shown in Figure~\ref{fig:optimization}(a), the input type conversion from |tensor<4x4x2x2xvector<8x8{>}{>}| to |itensor<16x16...>| indicates a DMA that will: 1) load |4x4x2x2| times |vector<8x8>| data from external memory; 2) store this data in a |16x16| ping-pong buffer to hide external memory access latency; and 3) push the data to a FIFO with a layout encoded in the |itensor| type. In Figure~\ref{fig:optimization}(b), we observe that \emph{DMA0} is automatically generated to implement these three behaviors using \texttt{scf.for} loop nests. Note that our |itensor|-based typing system encodes all the converter and DMA information. This is a capability that traditional tensor types lack, limiting their utility in dataflow component generation.



\subsubsection{Iterative Tensor Folding}

Figures~\ref{fig:optimization}(b)-(c) show the |itensor| folding. Suppose we have an |itensor\_write| in \emph{DMA0} and an |itensor\_read| in \emph{Kernel0}, connected via a FIFO. These represent two separate local buffers connected by streaming. By folding, we eliminate the FIFO and merge the two buffers. This optimization can reduce on-chip memory utilization while improving the overall latency by increasing the overlap between kernels. As shown in Figure~\ref{fig:optimization}(c), the fetched tile is directly passed to the |linalg.generic| op in \emph{Kernel0}, eliminating redundant buffering and communication. |itensor| folding requires an exact match in memory access patterns between producer and consumer. This makes it more restrictive than stream-based kernel fusion, which can be applied between any dataflow kernels. Consequently, we implement |itensor| folding as an additional optimization upon already fused kernels.

\subsubsection{Iterative Tensor Vectorization}

As dataflow kernels often run in parallel, we must vectorize dataflow FIFOs to provide sufficient bandwidth. Figures~\ref{fig:optimization}(c)-(d) show the vectorization of an |itensor| into |vector<2x4>|. On the \emph{DMA0+Kernel0} side, the |itensor\_write| becomes a loop with |transfer\_read| (from the buffer) followed by |itensor\_write| (to the FIFO). On the \emph{Converter0} side, similar transformations are applied for reading. This process aligns FIFO bandwidth with the parallelism of the dataflow kernel.

\section{Design Spaces}

To generate realizable and optimized accelerators, we must configure the compilation pass parameters properly. As shown in Figure~\ref{fig:framework}, we divide the overall design space into three sub-spaces: Linalg tiling space, kernel fusion space, and resource allocation space.


\subsection{Linalg Tiling Space}
\label{sec:linalg_tiling_space}

The Linalg tiling space determines tiling factors, unrolling factors, permutation strategies, and input/output vectorization for each dataflow kernel. In StreamTensor, this space is represented by a graph of Linalg operations, with properties such as loop trip counts, step sizes, and loop types (reduction or parallel) annotated on each node. The results of the exploration are also written back to this graph to configure transformation passes.

For tiling, a hyperparameter |default\_tile\_size| is exposed to users and applied across all dimensions of all kernels. For unrolling, we develop an intensity-aware algorithm, which iteratively selects the kernel with the longest latency through a max-heap and increases its unroll factor until a user-defined hyperparameter |overall\_unroll\_size| is reached. This approach balances kernel latencies to improve throughput. Once unroll sizes are determined, vectorization factors are inferred by analyzing the loop iteration space and tensor shapes. Permutation is handled by a heuristic that moves reduction loops outward while keeping parallel loops innermost, reducing initiation intervals (II) of pipeline loops. In StreamTensor, the hyperparameters of the Linalg tiling space are automatically explored through a blackbox optimizer, Optuna~\cite{akiba2019optuna}, with the feedback from dataflow kernel fusion results.

\subsection{Kernel Fusion Space}
\label{sec:kernel_fusion_space}

As described in Section~\ref{sec:fusion}, kernel fusion enables streaming between kernels. If the producer and consumer have different |itensor| types, a converter must be inserted. The exploration of the Linalg tiling space determines all data layouts and shapes, thereby fixing the |itensor| types at the interfaces of all dataflow kernels. Consequently, the memory overhead of fusing any pair of kernels is also established. Due to limited on-chip memory, fusing all kernels is generally not feasible. To effectively select which kernel pairs to fuse while adhering to memory resource constraints, we propose two algorithms: Algorithm~\ref{alg:get_converter_info} that infers the minimal ping-pong buffer shape required by the stream layout converter; and Algorithm~\ref{alg:greedy_kernel_fusion} that determines a global fusion plan under on-chip memory constraints.

\begin{algorithm}[t]
\small
\captionsetup{font=small}
\caption{Pseudo code of stream layout converter generation}
\label{alg:get_converter_info}
\begin{algorithmic}[1]
\Require $src$, Source |itensor| type; $res$, Result |itensor| type
\Ensure $bufShape$: Shape of the ping-pong buffer
\Ensure $beforeLoop$: Loop index where the ping-pong buffer is inserted

\State $bufShape \gets []$, $beforeLoop \gets 0$
\State $sharedLoops \gets []$ \Comment{Indices of loops shared by $src$ and $res$}
\For{$dim \gets 0$ \textbf{to} $src.rank() - 1$}
\If{$src.elementSize(dim) \neq res.elementSize(dim)$} \textbf{break}
\EndIf
\State $srcExpr \gets src.iterMap[dim]$
\State $resExpr \gets res.iterMap[dim]$

\If{both $Expr$s are dimensions with same position}
\State $bufShape.append(src.elementSize(dim))$
\State $sharedLoops.append(srcExpr.pos)$
\State $beforeLoop \gets beforeLoop + 1$
\Else ~\textbf{break}
\EndIf
\EndFor

\While{any $loop \in sharedLoops$ where $loop \geq beforeLoop$}
\State $bufShape.pop()$, $loop \gets sharedLoops.pop()$
\If{$loop \neq -1$} $beforeLoop \gets beforeLoop - 1$ \EndIf
\EndWhile

\State $bufShape.append(src.shape[bufShape.size():])$
\State \Return \{$bufShape$, $beforeLoop$\}
\end{algorithmic}
\end{algorithm}

\subsubsection{Stream Layout Converter Generation}
\label{sec:converter_generation}

Algorithm~\ref{alg:get_converter_info} compares the source and target |itensor|s across each data dimension (lines 3–16). The ping-pong buffer size can be reduced along a data dimension only if:
1) their element sizes are equal (lines 4-5); and 2) their corresponding iteration dimensions are equal, referring to the same loop nesting level (lines 8-16). For instance, in Figure~\ref{fig:itensor}, the second data dimensions of |itensor(b)| and |itensor(c)| both correspond to iteration dimension |d0|, allowing this dimension to be reduced; we only need to buffer a single column of tiles. In materialization, shared loops will be generated to reuse the buffer along this reduced dimension. Conversely, their first data dimensions correspond to iteration dimensions |d1| and |d2|, respectively, making them non-reducible. Thus, we must buffer all rows of tiles. Consequently, as Figure~\ref{fig:itensor} illustrates, two tiles (four tiles after ping-pong buffering) are required in the layout converter.

After identifying reducible data dimensions and corresponding shared loops, the algorithm filters out those that have parent loops that are not shareable, ensuring buffer realizability (lines 17–19). For example, if loop-|\{0,1,2,4\}| are shareable but loop-3 is not, loop-4 must be excluded. Finally, the buffer shape and shared loops are returned. This process's worst case occurs when no dimension is reducible, demanding that the entire data be held on-chip for fusion. This may result in significant memory overhead.

\begin{algorithm}[t]
\small
\captionsetup{font=small}
\caption{Pseudo code of kernel fusion exploration}
\label{alg:greedy_kernel_fusion}
\begin{algorithmic}[1]
\Require $G$, kernel fusion design space; $C_{max}$, max fusion cost
\Ensure $F$, sets of nodes to be fused; $C$, costs of fused nodes

\State $F \gets [\emptyset]$, $C \gets [0]$, $M \gets \{\}$ \Comment{Map from node to index of fusion}
\For{$n$ \textbf{in} $topo\_sort(G)$}

\State $cand \gets \{\}$ \Comment{Map from index of fusion candidate to cost}
\For{$p$ \textbf{in} $G.predecessors(n)$}
\State $cost \gets$ compute\_memory\_cost($G.edges[p,n,0]$)
\State $cand[M[p]] \gets cand.get(M[p], 0) + cost$
\EndFor

\State $f\_idx \gets len(F)$, $f\_cost \gets 0$
\If{$len(cand) > 0$} \Comment{Fuse with the nearest candidate}
\State $f\_idx \gets \max(cand.keys())$, $f\_cost \gets cand[f\_idx]$
\EndIf

\If{$f\_idx = len(F)$ \textbf{or} $f\_cost + C[f\_idx] > C_{max}$}
\State $F.append(\{n\})$, $C.append(0)$, $M[n] \gets len(F) - 1$
\Else
\State $F[f\_idx].add(n)$, $C[f\_idx] \gets C[f\_idx]+f\_cost$
\State $M[n] \gets f\_idx$
\EndIf
\State $G.nodes[n]["fusion\_index"] \gets M[n]$
\EndFor
\State \Return $F$, $C$
\end{algorithmic}
\end{algorithm}

\begin{figure*}
    \centering
    \includegraphics[width=\linewidth]{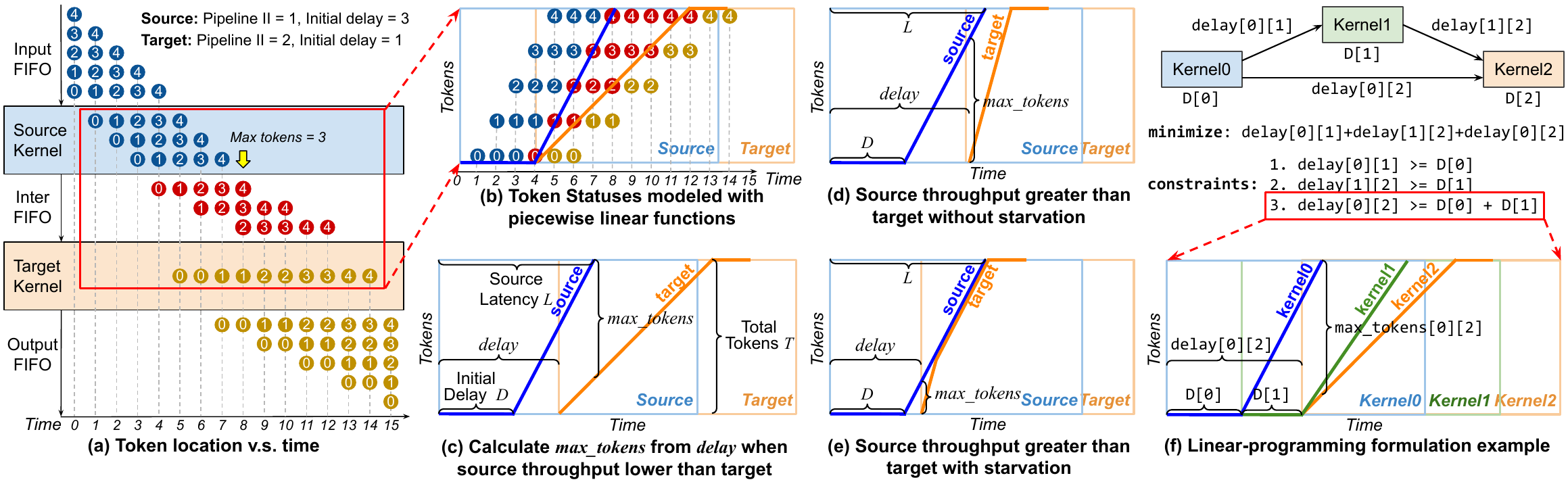}
    \caption{Token behavior modeling with piecewise linear function and linear-programming-based FIFO sizing formulation.}
    \label{fig:sizing_strategy}
\end{figure*}

\subsubsection{Kernel Fusion Exploration}

The input $C_{max}$ (\emph{max fusion cost}) for Algorithm~\ref{alg:greedy_kernel_fusion} represents the maximum on-chip memory a single fused kernel can utilize. For FPGAs, this is typically set to the total on-chip memory size. Consequently, the kernel fusion process can also be viewed as a graph partitioning problem. After fusion, each resulting fused kernel will occupy a single FPGA. If a computation graph comprises multiple such kernels, they can be executed across multiple FPGAs, on a single FPGA sequentially, or with a hybrid approach. StreamTensor supports all these approaches as a compiler. However, mapping $M$ kernels to $N$ FPGAs and managing inter-FPGA communication are beyond the scope of this paper. Algorithm~\ref{alg:greedy_kernel_fusion} traverses all |kernel|s in a topological order (line 3). For each kernel, it first gathers fusion candidates from predecessors and computes the fusion cost (lines 4–11). The kernel is fused with the nearest valid candidate (lines 13–14) if it does not exceed the resource limit (lines 15–20). Fusion results are written back to the graph (line 22) and used to configure the optimizations discussed in Section~\ref{sec:fusion}. Dataflow kernel fusion always has a feasible solution unless a single kernel occupies more resources than a single FPGA. In such a case, the result is fed back to the tiling space for refinement, for example, reducing tiling and/or unrolling factors.

\subsection{Resource Allocation Space}
\label{sec:resource_allocation}

On hardware like FPGAs, due to limited on-chip memory and compute resources, effective resource allocation greatly affects routing congestion and clock frequency. In this space, we need to solve:
\begin{enumerate}[leftmargin=*]
    \item \textbf{FIFO sizing}: Determine FIFO depths to avoid deadlocks and improve execution overlap. This section will cover more details.
    \item \textbf{Graph partitioning}: On multi-die hardware, we need to assign |task|s to dies. This assignment problem is formulated and solved using Integer Linear Programming (ILP). In our ILP model, a binary list represents each |task|'s assignment. A constraint ensures that only one element in this list can be "1", with its position indicating the assigned die. The ILP objective is to minimize both inter-die communication and resource imbalance across the dies. Since similar formulations have been studied~\cite{guo2021autobridge,du2023fado}, we omit further details.
    \item \textbf{Memory allocation}: Place each buffer in LUTRAM, BRAM, or URAM on FPGAs, prioritized by size. Since this algorithm is straightforward, we omit further details.
\end{enumerate}

\subsubsection{Token Behavior Model}
\label{sec:token_model}

To address the FIFO sizing problem discussed in Section~\ref{sec:pitfalls}, we first propose a token production and consumption model based on piecewise linear functions. Figure~\ref{fig:sizing_strategy}(a) illustrates the token communication between \emph{Source} and \emph{Target} kernels fused through \emph{InterFIFO}. Pipeline II is the cycle count between two consecutive output tokens, while initial delay is the cycle count required to produce the first output token. A token is defined as the atomic data element communicated between kernels. At \emph{time0}, all five input tokens are in \emph{InputFIFO}, and tokens begin to stream into \emph{Source} at \emph{time1}. At \emph{time5}, \emph{Source} pushes \emph{token1} into \emph{InterFIFO}, while \emph{Target} consumes \emph{token0}, leaving one token in \emph{InterFIFO}. At \emph{time6}, \emph{Target} cannot consume \emph{token1} because it requires two cycles to process \emph{token0}. Meanwhile, \emph{token2} is pushed into \emph{InterFIFO}, increasing its token count to two. At \emph{time8}, \emph{Source} finishes processing tokens, when \emph{InterFIFO} holds its maximum capacity of three tokens.
\emph{Target} then continues to consume and process the remaining tokens until \emph{time15}, when all tokens are fully processed.

To model these complex behaviors with an analyzable function, we reorganize the token statuses from Figure~\ref{fig:sizing_strategy}(a) into Figure~\ref{fig:sizing_strategy}(b), aligning the statuses of the same token in the same row. We observe that the boundary between the \emph{Source}~(blue) and \emph{InterFIFO}~(red) sections can be perfectly modeled with a piecewise linear function~(blue curve). This function represents the token count \emph{produced} by \emph{Source}. Similarly, we can model the token count \emph{consumed} by \emph{Target} with the orange curve. The difference between these two curves represents the token count in \emph{InterFIFO}. These curves can be represented by the kernel's latency, initial delay, and pipeline II. StreamTensor automatically invokes vendor tools like HLS to profile these metrics for each kernel in the middle of the flow. Since these metrics are specific to vendor platform's architecture, technology node, and mapping strategy, they must be obtained through this profiling process. As resource allocation is the last design space, the kernel designs remain unchanged in the subsequent StreamTensor flow. As long as the vendor tools use a deterministic scheduling algorithm, the final accelerator's metrics will match those profiled earlier. This consistency guarantees the validity of our algorithm.

\subsubsection{Maximum Token Calculation}
\label{sec:max_tokens}

As shown in Figure~\ref{fig:sizing_strategy}(c), we define $L$ as the total latency of \emph{Source} execution; $D$ as the initial delay from the start of \emph{Source} execution to the production of its first output token; $delay$ as the time from the start of \emph{Source} execution to the start of \emph{Target} execution. Naturally, $delay$ is always greater than or equal to $D$ since \emph{Target} cannot start its execution before the first token is produced by \emph{Source}. We define $T$ as the exact number of tokens passed from \emph{Source} to \emph{Target} for a single accelerator execution. $T$ is a static value that can be analytically inferred from tensor shapes in StreamTensor. We will address how to handle dynamic tensor shapes in Section~\ref{sec:dynamic_behaviors}. With a static $T$ value, the maximum token count in \emph{InterFIFO}, $max\_tokens$, can be analytically calculated from $delay$:
\begin{equation}
\label{eq:larger_source}
    max\_tokens = \min\left(T,~T - \lfloor\frac{L - delay}{II_{Target}}\rfloor\right)
\end{equation}

The pipeline $II$ determines the slope of the curve, i.e., the kernel throughput. Figure~\ref{fig:sizing_strategy}(c) illustrates the case where \emph{Source}'s throughput is greater than \emph{Target}'s. Conversely, when \emph{Source}'s throughput is lower, data starvation may limit \emph{Target}'s throughput. Figure~\ref{fig:sizing_strategy}(d) shows that \emph{Target} is unaffected with a sufficiently large $delay$, whereas Figure~\ref{fig:sizing_strategy}(e) shows that \emph{Target} is eventually starved and its throughput is equalized to \emph{Source}'s throughput. In both cases, $max\_tokens$ can be calculated from $delay$:
\begin{equation}
\label{eq:smaller_source}
    max\_tokens = \min\left(T,~\lceil\frac{delay - D}{II_{Source}}\rceil\right)
\end{equation}

Equations~\ref{eq:larger_source} and \ref{eq:smaller_source} both reveal a positive correlation between $max\_tokens$ and $delay$. As shown in Figure~\ref{fig:sizing_strategy}(c)-(e), setting the \emph{InterFIFO} depth to $max\_tokens$ prevents back-pressure from \emph{Target} onto \emph{Source}. This ensures steady, periodic behavior between any pair of \emph{Source} and \emph{Target} across multiple accelerator executions. By preventing stalls from back-pressure, the analytical relationship between $max\_tokens$ and $delay$ is preserved.

\subsubsection{Equalization}
\label{sec:equalization}

The approach described in Section~\ref{sec:max_tokens} is named as the \emph{Normal} equalization strategy, which assumes that kernels always produce tokens at their original throughput. However, the throughput of a dataflow accelerator is ultimately determined by its slowest kernel. Based on this, we propose a \emph{Conservative} equalization strategy, which \emph{scales} the pipeline II of all kernels to match the throughput of the slowest kernel. The resulting \emph{max\_tokens} values are smaller than or equal to those from the \emph{Normal} strategy because the gap between any pair of \emph{Source} and \emph{Target} curves is minimized. The drawback is that faster kernels are frequently stalled by back-pressure, potentially increasing the latency. Therefore, the \emph{Normal} and \emph{Conservative} strategies present a trade-off between area and performance, where the \emph{Conservative} strategy minimizes FIFO buffer sizes at the cost of increased overall latency. The key difference between the \emph{Conservative} and \emph{Normal} strategies lies in how their IIs are initially scaled.
Because this scaling preserves the piecewise-linear nature of the kernel curves, the equations for calculating $max\_tokens$ from $delay$ remain identical for both strategies.

\subsubsection{LP-based FIFO Sizing}

By introducing the token behavior model, we transform the FIFO sizing problem into a problem of determining the $delay$ values between kernels. Figure~\ref{fig:sizing_strategy}(f) shows an example of dataflow graph. \emph{Kernel0} has two outputs; \emph{Kernel1} depends on \emph{Kernel0}; \emph{Kernel2} has two operands and must wait for both \emph{Kernel0}'s and \emph{Kernel1}'s first tokens. Given that \emph{Kernel1} produces its first token after |D[0]+D[1]|, |delay[0][2]| must be greater than or equal to this value. Their relationship is depicted in Figure~\ref{fig:sizing_strategy}(f), with the green curve representing \emph{Kernel1}. The maximum token count for the FIFO between \emph{Kernel0} and \emph{Kernel2}, |max\_token[0][2]|, can then be calculated using |delay[0][2]|. If the FIFO size is smaller than this maximum, \emph{Kernel0} will stall due to back-pressure, which harms overall performance. This stall can propagate to \emph{Kernel1} and \emph{Kernel2}, preventing the back-pressure from resolving and potentially causing a deadlock. A FIFO size equal to $|max\_token[0][2]|$ is sufficient to prevent back-pressure and avoid a deadlock; it is also required to prevent performance degradation from unintended kernel stalls. We propose an LP formulation to optimally solve for the $delay$ values. Given $G = (V, E)$, where $V$ is the set of kernels and $E$ is the set of edges between the kernels, the objective and constraints of LP are:
\begin{align}
    minimize & \sum_{e_{i,j} \in E} delay(i,j) \\
    \forall u, v \in V, \forall path \in P_{u,v}, & \sum_{e_{i,j} \in path} delay(i, j) \geq threshold(u, v)
\end{align}

$e_{i,j} \in E$ covers all edges in the graph; $path \in P_{u,v}$ covers all full paths connecting any pair of kernels, named |u| and |v|; $e_{i,j} \in path$ covers all edges along a $path$ connecting the two kernels |u| and |v|. We minimize the summation of |delay|s on all edges, which serves as a proxy for optimizing FIFO sizes due to the positive correlation between $max\_tokens$ and $delay$. $threshold(u, v)$ is the maximum accumulated $D$ over all paths connecting the two kernels |u|~and~|v|:
\begin{align}
    threshold(u, v) = \max_{path \in P_{u,v}} \sum_{e_{i,j} \in path} D(i)
\end{align}

The LP formulation for the example above is shown in Figure~\ref{fig:sizing_strategy}(f). Note that in this example, the two paths diverging from \emph{Kernel0} re-converge to \emph{Kernel2} as two distinct input operands, rather than joining into a single input. We will discuss the handling of dynamic behaviors like path joining in Section~\ref{sec:dynamic_behaviors}. Resource constraints are not needed for the LP problem for two reasons: First, as discussed in Section~\ref{sec:kernel_fusion_space}, dataflow kernel fusion guarantees that all fused kernels will fit within available on-chip resources by restricting the fusion cost. Second, the memory utilization of stream FIFOs is negligible compared to that of dataflow kernels and converters. Consequently, the LP problem can be optimally solved in polynomial time. 
Notably, we do not need to enforce vendor tools to implement the $delay$s. Instead, the $delay$s are automatically fulfilled through the FIFO dependencies between dataflow kernels.
In the example above, \emph{Kernel2} automatically waits for \emph{Kernel1} because it depends on \emph{Kernel1}'s output token.

\begin{table*}
    \centering
    \small
    \setlength\tabcolsep{5pt}
    \caption{Comparison with previous works on GPT-2 model. \emph{TTFT} measures the time to first token in ms, the lower the better. \emph{Speed} measures the decoding speed in token/s, the higher the better. All results of previous works are directly from their papers.}
    \label{tab:comparison_gpt2}
    \begin{tabular}{cccccccccc}
        \toprule
        \multirow{2}{*}[-7pt]{\multirowcell{c}{\textbf{[Input Len:}\\\textbf{Output Len]}}} & \multicolumn{3}{c}{\textbf{Ours}} & \multicolumn{3}{c}{\textbf{Allo~\cite{chen2024allo} (Ratio of $\frac{Ours}{Allo}$)}} & \multicolumn{3}{c}{\textbf{DFX~\cite{hong2022dfx} (Ratio of $\frac{Ours}{DFX}$)}} \\
         \cmidrule(lr){2-4} \cmidrule(lr){5-7} \cmidrule(lr){8-10}
         & \multirowcell{c}{\textbf{Latency} \\ \textbf{(ms)}} & \multirowcell{c}{\textbf{TTFT} \\ \textbf{(ms)}} & \multirowcell{c}{\textbf{Speed} \\ \textbf{(token/s)}} & \multirowcell{c}{\textbf{Latency} \\ \textbf{(ms)}} & \multirowcell{c}{\textbf{TTFT} \\ \textbf{(ms)}} & \multirowcell{c}{\textbf{Speed} \\ \textbf{(token/s)}} & \multirowcell{c}{\textbf{Latency} \\ \textbf{(ms)}} & \multirowcell{c}{\textbf{TTFT} \\ \textbf{(ms)}} & \multirowcell{c}{\textbf{Speed} \\ \textbf{(token/s)}} \\
        \midrule
        \textbf{[32:32]} & 194.99 & 34.59 & 199.51 & 238.32 (0.82x) & 81.50 (0.42x) & 204.05 (0.98x) & 350.00 (0.56x) & 177.20 (0.20x) & 185.19 (1.08x) \\
        \textbf{[64:64]} & 358.24 & 61.27 & 215.51 & 476.64 (0.75x) & 162.99 (0.38x) & 204.05 (1.06x) & 694.70 (0.52x) & 349.10 (0.18x) & 185.19 (1.16x) \\
        \textbf{[128:128]} & 696.65 & 125.35 & 224.05 & 953.28 (0.73x) & 325.98 (0.38x) & 204.05 (1.10x) & 1384.00 (0.50x) & 692.80 (0.18x) & 185.19 (1.21x) \\
        \textbf{[256:256]} & 1387.76 & 272.85 & 229.61 & 1906.56 (0.73x) & 651.96 (0.42x) & 204.05 (1.13x) & 2800.00 (0.50x) & 1417.60 (0.19x) & 185.19 (1.24x) \\
        \midrule
        \textbf{Geo. Mean} & - & - & - & \textbf{0.76x} & \textbf{0.40x} & \textbf{1.06x} & \textbf{0.52x} & \textbf{0.19x} & \textbf{1.17x} \\
        \bottomrule
    \end{tabular}
\end{table*}

\begin{table*}[]
    \centering
    \small
    \setlength\tabcolsep{5pt}
    \caption{Comparison with NVIDIA GPUs on GPT-2 model. \emph{TTFT} measures the time to first token in ms, the lower the better. \emph{Speed} measures the decoding speed in token/s, the higher the better.}
    \label{tab:comparison_gpu_gpt2}
    \begin{tabular}{ccccccccccc}
        \toprule
        \multirow{2}{*}[-7pt]{\multirowcell{c}{\textbf{[Input Len:}\\\textbf{Output Len]}}} & \multicolumn{3}{c}{\textbf{Ours}} & \multicolumn{3}{c}{\textbf{A100 (Ratio of $\frac{Ours}{A100}$)}} & \multicolumn{3}{c}{\textbf{2080Ti (Ratio of $\frac{Ours}{2080Ti}$)}} \\
        \cmidrule(lr){2-4} \cmidrule(lr){5-7} \cmidrule(lr){8-10}
         & \multirowcell{c}{\textbf{Latency} \\ \textbf{(ms)}} & \multirowcell{c}{\textbf{TTFT} \\ \textbf{(ms)}} & \multirowcell{c}{\textbf{Speed} \\ \textbf{(token/s)}} & \multirowcell{c}{\textbf{Latency} \\ \textbf{(ms)}} & \multirowcell{c}{\textbf{TTFT} \\ \textbf{(ms)}} & \multirowcell{c}{\textbf{Speed} \\ \textbf{(token/s)}} & \multirowcell{c}{\textbf{Latency} \\ \textbf{(ms)}} & \multirowcell{c}{\textbf{TTFT} \\ \textbf{(ms)}} & \multirowcell{c}{\textbf{Speed} \\ \textbf{(token/s)}} \\
        \midrule
        \textbf{[32:32]} & 194.99 & 34.59 & 199.51 & 291.16 (0.67x) & 8.72 (3.97x) & 113.30 (1.76x) & 518.46 (0.38x) & 24.98 (1.38x) & 64.85 (3.08x) \\
        \textbf{[64:64]} & 358.24 & 61.27 & 215.51 & 567.41 (0.63x) & 8.76 (6.99x) & 114.56 (1.88x) & 1010.81 (0.35x) & 25.23 (2.43x) & 64.94 (3.32x) \\
        \textbf{[128:128]} & 696.65 & 125.35 & 224.05 & 1118.28 (0.62x) & 8.65 (14.49x) & 115.35 (1.94x) & 3969.76 (0.18x) & 25.26 (4.96x) & 32.45 (6.90x) \\
        \textbf{[256:256]} & 1387.76 & 272.85 & 229.61 & 2227.79 (0.62x) & 8.53 (31.99x) & 115.35 (1.99x) & 7914.23 (0.18x) & 25.23 (10.81x) & 32.45 (7.08x) \\
        \midrule
        \textbf{Geo. Mean} & - & - & - & \textbf{0.64x} & 10.65x & \textbf{1.89x} & \textbf{0.25x} & 3.67x & \textbf{4.73x} \\
        \bottomrule
    \end{tabular}
\end{table*}

\subsubsection{Dynamic Behaviors}
\label{sec:dynamic_behaviors}

StreamTensor uses different approaches to manage dynamic behaviors within dataflow accelerators:
\begin{enumerate}[leftmargin=*]
    \item \textbf{Control flow}: StreamTensor leverages Torch-MLIR~\cite{torch_mlir_github} as its front-end. Torch-MLIR can infer the static tensor shapes as much as possible from inputs, eliminating \texttt{if}s and unrolling \texttt{for}s associated with static tensor shapes. If the control flow relies on runtime values, the corresponding subgraph will fall back to naive PyTorch execution~\cite{ansel2024pytorch} on the host.
    \item \textbf{Path joining}: This often arises in the presence of control flow, particularly when a dataflow kernel is reused with inputs from different sources. By eliminating control flows, Torch-MLIR resolves the corresponding path joining problems.
    \item \textbf{Dynamic tensor shape}: Tensors with dynamic shapes, like input tokens and KV-caches, require shape hints to define their maximum possible dimension sizes (e.g., maximum sequence length). These hints determine the total number of tokens, $T$, that can be processed between any two dataflow kernels. From these maximum $T$ values, StreamTensor infers $max\_tokens$ based on the method discussed in Section~\ref{sec:resource_allocation}.
    \item \textbf{FIFO stall}: StreamTensor does not generate a static schedule for the dataflow accelerator. Instead, all dataflow kernels automatically honor their dependencies via FIFO interconnections. As a result, unexpected FIFO stalls caused by runtime events, e.g., external memory traffic, do not require specific handling. Once the event causing the stall resolves, the dataflow accelerator seamlessly resumes operation from the stall point.
\end{enumerate}

\section{Experiments}

To evaluate the performance of dataflow accelerators generated by StreamTensor, we deploy multiple LLMs on AMD U55C FPGA with Vitis~2024.1. As shown in Figure~\ref{fig:framework}, HLS C++ code is generated by StreamTensor and compiled into bitstreams using Vitis to program the FPGA.
Table~\ref{tab:exp_setup} shows the experimental setup of the platforms evaluated in this section.
All experimental results of StreamTensor reported are obtained via \emph{on-board measurement}. All LLM models evaluated on StreamTensor are modified from Huggingface models to accommodate the requirements of Torch-MLIR front-end.

\begin{table}[t]
    \centering
    \small
    \setlength\tabcolsep{3pt}
    \caption{Experiment setup of evaluated platforms.}
    \label{tab:exp_setup}
    \begin{tabular}{cccccc}
        \toprule
         & \textbf{Ours} & \textbf{Allo~\cite{chen2024allo}} & \textbf{DFX~\cite{hong2022dfx}} & \textbf{A100} & \textbf{2080Ti} \\
        \midrule
        \textbf{Platform} & \multirowcell{c}{AMD \\ U55C} & \multirowcell{c}{AMD \\ U280} & \multirowcell{c}{AMD \\ U280} & \multirowcell{c}{NVIDIA \\ A100} & \multirowcell{c}{NVIDIA \\ 2080Ti} \\
        \hline
        \textbf{Process Node} & 16nm & 16nm & 16nm & 7nm & 12nm \\
        \hline
        \textbf{Freq. (MHz)} & 250 & 250 & 200 & 1065 & 1350 \\
        \hline
        \textbf{Quantization} & W4A8 & W4A8 & FP16 & W8A8 & W8A8 \\
        \hline
        \multirowcell{c}{\textbf{Thermal} \\ \textbf{Design Power}} & 150W & 225W & 225W & 300W & 250W \\
        \hline
        \multirowcell{c}{\textbf{Peak Perf.} \\ \textbf{(INT8 TOPS)}} & 24.5 & 24.5 & 24.5 & 624 & 215.2 \\
        \hline
        \multirowcell{c}{\textbf{Off-chip} \\ \textbf{Memory}} & \multirowcell{c}{16GB \\ HBM \\ 460GB/s} & \multirowcell{c}{8GB \\ HBM \\ 460GB/s} & \multirowcell{c}{8GB \\ HBM \\ 460GB/s} & \multirowcell{c}{80GB \\ HBM \\ 1935GB/s} & \multirowcell{c}{11GB \\ DDR \\ 616GB/s} \\
        \hline
        \multirowcell{c}{\textbf{On-chip} \\ \textbf{Memory}} & 41MB & 41MB & 41MB & 40MB & 5.5MB \\
        \bottomrule
    \end{tabular}
\end{table}

\begin{figure*}
    \centering
    \begin{subfigure}[t]{0.34\textwidth}
        \centering
        \includegraphics[width=0.98\linewidth]{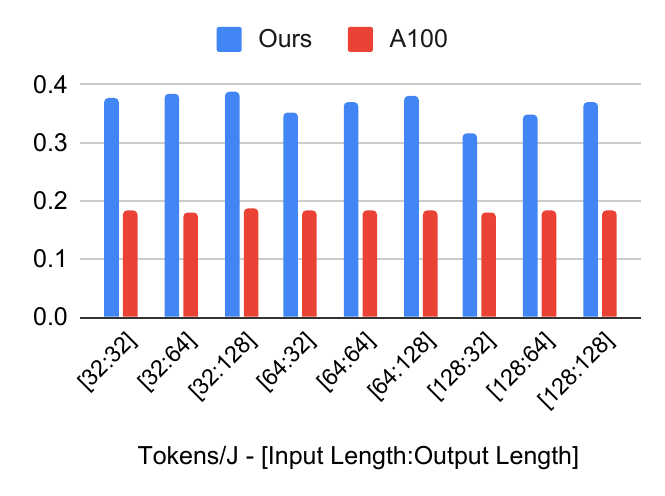}
        \caption{Qwen.}
        \label{fig:efficiency_qwen}
    \end{subfigure}
    \hspace{-0.04\textwidth}
    \begin{subfigure}[t]{0.34\textwidth}
        \centering
        \includegraphics[width=0.98\linewidth]{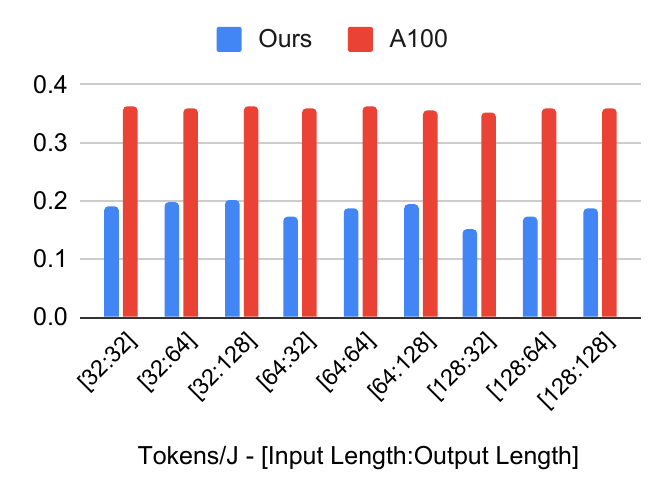}
        \caption{Llama.}
        \label{fig:efficiency_llama}
    \end{subfigure}
    \hspace{-0.04\textwidth}
    \begin{subfigure}[t]{0.34\textwidth}
        \centering
        \includegraphics[width=0.98\linewidth]{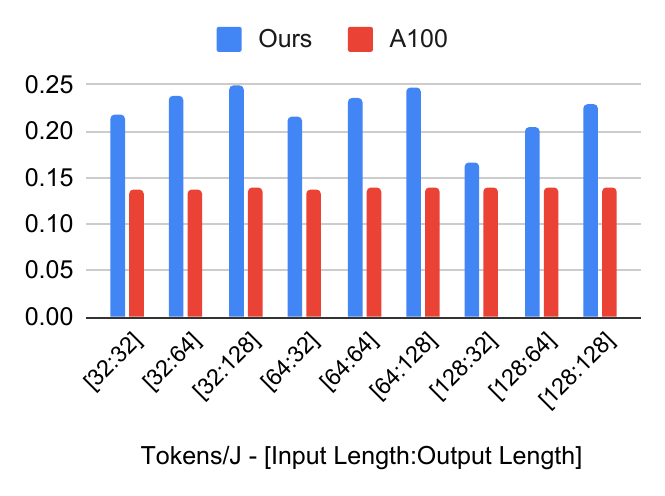}
        \caption{Gemma.}
        \label{fig:efficiency_gemma}
    \end{subfigure}
    \caption{Energy efficiency (tokens/J) comparison with NVIDIA GPUs on emerging LLMs.}
    \label{fig:comparison_llm_efficiency}
\end{figure*}

\begin{figure*}
    \centering
    \begin{subfigure}[t]{0.34\textwidth}
        \centering
        \includegraphics[width=0.98\linewidth]{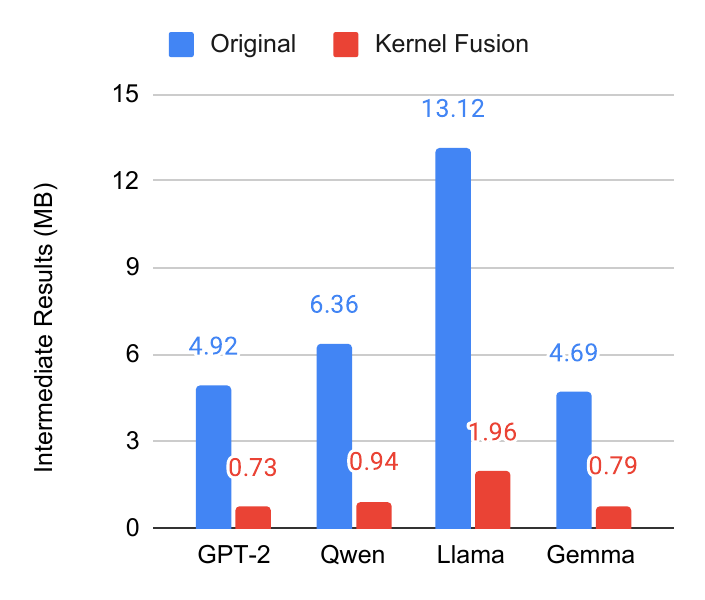}
        \caption{Memory reduction.}
        \label{fig:memory_reduction}
    \end{subfigure}
    \hspace{-0.03\textwidth}
    \begin{subfigure}[t]{0.34\textwidth}
        \centering
        \includegraphics[width=0.98\linewidth]{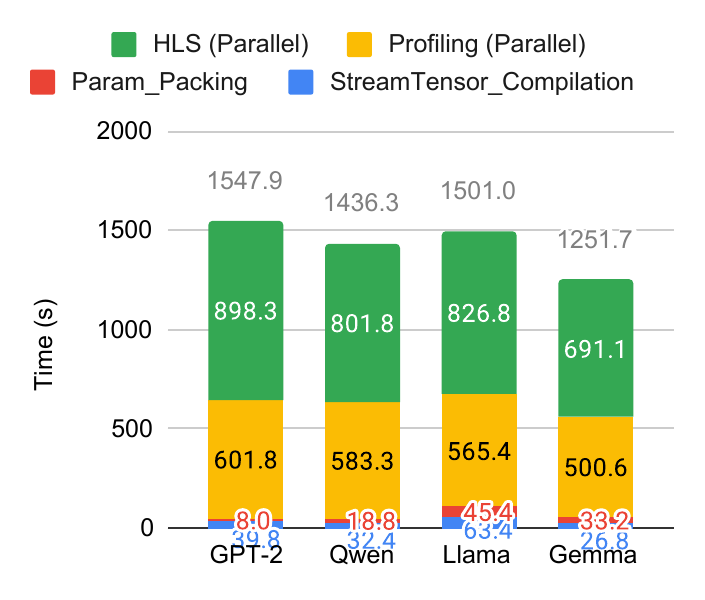}
        \caption{RTL generation time (s).}
        \label{fig:overall_time}
    \end{subfigure}
    \hspace{-0.03\textwidth}
    \begin{subfigure}[t]{0.34\textwidth}
        \centering
        \includegraphics[width=0.98\linewidth]{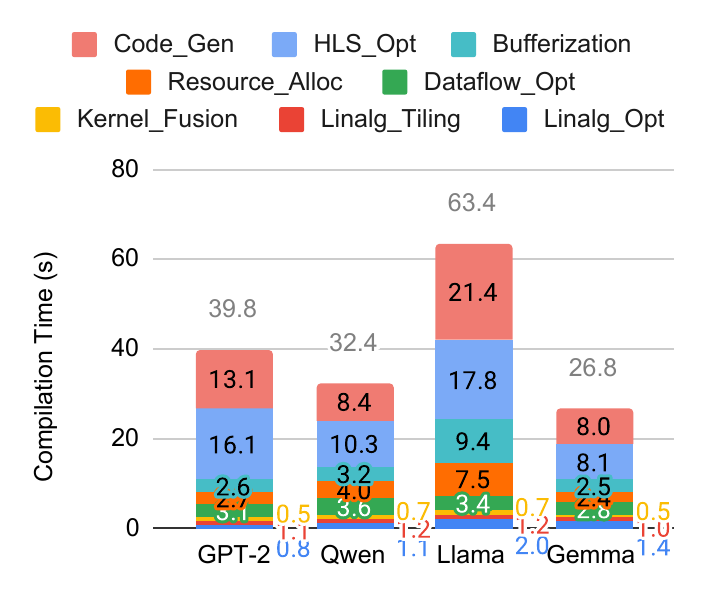}
        \caption{StreamTensor compile time (s).}
        \label{fig:compile_time}
    \end{subfigure}
    \caption{Ablation studies on GPT-2 model and emerging LLMs.}
\end{figure*}

\subsection{GPT-2}

Most prior works~\cite{chen2024allo,chen2024understanding,hong2022dfx} on FPGAs evaluate their frameworks using GPT-2~\cite{radford2019language}. Table~\ref{tab:comparison_gpt2} shows a comparison between StreamTensor and previous works under different input/output sequence length configurations. For GPT-2, we successfully fuse an entire transformer block onto a single FPGA by inserting layout converters and stream FIFOs, ensuring all intermediate results are communicated on-chip. Subsequently, this single FPGA accelerator is triggered multiple times with different weight parameters to execute all transformer blocks in a sequential manner. StreamTensor achieves 0.76x shorter total latency and 0.40x shorter TTFT than Allo~\cite{chen2024allo,chen2024understanding}. Compared to DFX~\cite{hong2022dfx}, StreamTensor delivers even greater improvements, e.g., 0.19x TTFT. These gains come from StreamTensor's automated dataflow architecture exploration. In contrast, both Allo and DFX require manual design of all dataflow kernels and components. For example, all the layout converters, DMAs, and FIFOs are manually written and configured, a process that is error-prone and may lead to suboptimal design choices. Note that GPT-2 is the only LLM reported in Allo and DFX due to their limited flexibility and productivity on other emerging LLMs. As shown in Table~\ref{tab:comparison_gpt2}, TTFT scales roughly linearly with input length, demonstrating the design's scalability. We also compare StreamTensor with NVIDIA GPUs in Table~\ref{tab:comparison_gpu_gpt2}, where StreamTensor achieves 0.64x and 0.25x shorter total latency compared to A100 and 2080Ti, respectively. We can observe that GPUs outperform StreamTensor by a large margin for the TTFT metric due to their abundant computation resources. However, because the decoding stage of LLM inference is highly memory-bound, the dataflow accelerators generated by StreamTensor can outperform GPUs due to their reduced external memory access, leading to better decoding speed and overall latency.

\begin{table}[t]
    \centering
    \small
    \setlength\tabcolsep{3pt}
    \caption{Configurations of LLMs, collected from their Huggingface model cards~\cite{gpt2_openai_community, qwen2_5_0_5b, llama_3_2_1b, gemma3_1b_it}.}
    \label{tab:llm_config}
    \begin{tabular}{ccccc}
        \toprule
         & \textbf{GPT-2~\cite{radford2019language}} & \textbf{Qwen~\cite{bai2023qwen}} & \textbf{Llama~\cite{touvron2023llama}} & \textbf{Gemma~\cite{team2024gemma}} \\
        \midrule
        \textbf{Layers} & 24 & 24 & 22 & 26 \\
        \hline
        \textbf{Hidden Size} & 1024 & 896 & 2048 & 1152 \\
        \hline
        \multirowcell{c}{\textbf{FFN} \\ \textbf{Hidden Size}} & 4096 & 4864 & 5632 & 6912 \\
        \hline
        \multirowcell{c}{\textbf{Attention} \\ \textbf{Heads}} & 16 & 14 & 32 & 4 \\
        \hline
        \textbf{KV Heads} & - & 2 & 4 & 1 \\
        \hline
        \textbf{Activation} & GELU & SiLU & SiLU & GELU \\
        \bottomrule
    \end{tabular}
\end{table}

\subsection{Emerging LLMs}

To evaluate the flexibility of StreamTensor, we test it on several emerging LLMs, including Qwen~\cite{bai2023qwen}, Llama~\cite{touvron2023llama}, and Gemma~\cite{team2024gemma}. Model configurations are shown in Table~\ref{tab:llm_config}. For all three of these models, we also successfully fuse an entire transformer block onto a single FPGA and execute it in the same manner as GPT-2. From Figure~\ref{fig:comparison_llm_efficiency}, we observe that StreamTensor can outperform A100 on energy efficiency on Qwen and Gemma models by 1.99x and 1.59x due to the lower power of FPGAs. Figure~\ref{fig:memory_reduction} shows that the Llama model generates more intermediate results than other models. This leads StreamTensor to adopt a more conservative dataflow FIFO sizing strategy, which, in turn, reduces the execution overlap between dataflow kernels and results in lower performance compared to Qwen and Gemma.


\subsubsection{On-chip Memory Reduction Study}

Figure~\ref{fig:memory_reduction} shows on-chip memory usage before and after kernel fusion across all evaluated LLMs. This study focuses on the \emph{intermediate results} within a single LLM layer. Model parameters are excluded in this study, as they are too large to fit on-chip. Kernel fusion reduces memory usage to just 14.8\%–16.8\% of the original design. Without fusion, LLMs cannot be deployed in a fully dataflow fashion due to excessive intermediate buffer sizes.

\subsubsection{Compilation Time Study}

Figure~\ref{fig:overall_time} shows the breakdown of execution time for generating RTL from PyTorch. The HLS process (generating RTL from C++) consumes the majority of the total time. The downstream tool profiling also accounts for a large portion, since resource allocation decisions depend on accurate profiling results. In comparison, StreamTensor compilation and parameter packing take only a small fraction of the total time. As discussed in Section~\ref{sec:fusion}, StreamTensor automatically packs and widens interfaces to optimize external memory efficiency. As a result, model parameters must be packed accordingly to match the desired memory layout. After packing, binary files are generated and loaded at runtime. In Figure~\ref{fig:compile_time}, we further break down StreamTensor’s compilation time based on the stages shown in Figure~\ref{fig:framework}. Total compilation time ranges from 26.8s to 63.4s in our experiments. High-level stages (from Linalg optimization to resource allocation) are relatively fast. In contrast, low-level stages (bufferization, HLS optimization, and code generation) take more time. This validates the efficiency of our high-level |itensor| optimizations.

\section{Related Works}

Pioneering works~\cite{lee1987synchronous,bilsen1996cycle,bhattacharya2001parameterized,thies2002streamit,neuendorffer2004hierarchical} established the foundation of stream-based dataflow modeling and compilation. Later works~\cite{govindarajan2002minimizing,venkataramani2006leveraging,najibi2013slack} explored buffer minimizing and slack matching problems in dataflow networks. \cite{cong2014combining,cheng2016synthesis}~explored the deadlock analysis and buffer sizing for sequential programs. Note that these papers focused on steady-state scenarios (i.e., the \emph{Conservative} equalization strategy in Section~\ref{sec:equalization}), overlooking the trade-off between area and performance. \cite{guo2021autobridge,du2023fado}~improved the floorplanning and clock frequency for streaming applications on FPGAs. \cite{josipovic2021buffer,xu2024suppressing} tackled the buffer insertion and placement problem in dynamically scheduled dataflow circuits~\cite{josipovic2018dynamically}.

Compilers are essential for mapping applications onto spatial architectures like DSAs and FPGAs. SARA~\cite{zhang2021sara} provided a compiler stack for large-scale DSAs like Plasticine~\cite{prabhakar2017plasticine}, translating an imperative DSL with nested control flow, virtualizing resources, and managing memory consistency. The compiler for Revet~\cite{rucker2023revet} mapped its “dataflow threads” abstraction, which supports data-dependent control flow, onto vectorized DSAs~\cite{rucker2021capstan} using streaming tensor operations. Works like DSAGEN~\cite{weng2020dsagen} synthesized programmable spatial accelerators directly from dataflow graph descriptions. Constraint-based scheduling techniques~\cite{nowatzki2013general} often use ILP for optimal or near-optimal instruction scheduling on spatial platforms. Higher-level programming abstractions are also crucial, such as Sigma~\cite{zhao2023sigma}, which compiled Einstein summations to dataflow hardware. Targeting FPGAs, Stream-HLS~\cite{basalama2025stream} automatically generated optimized HLS-based dataflow architectures from C/C++ or PyTorch. These diverse compilers and frameworks automated critical optimizations. However, they often only enable partial design space exploration, and lack a systematic typing system to enable flexible stream-based kernel fusion and other optimizations. Here, we use Stream-HLS~\cite{basalama2025stream} as an example to analyze its differences with StreamTensor:
\begin{itemize}[leftmargin=*]
\item Due to the lack of a systematic typing system, Stream-HLS cannot automatically generate DMAs for external memory, limiting its practical usage and scalability on real-world applications.
\item Stream-HLS overlooked the FIFO sizing problem, which is essential to avoid deadlocks in dataflow accelerators and scale out to real-world applications.
\item Stream-HLS demanded two conditions to enable streaming between dataflow kernels: 1) the number of writes and reads to/from the shared buffer must be equal; and 2) the write order of the producer must match the read order of the consumer. Although both conditions are often difficult to meet, Stream-HLS cannot perform kernel fusion without meeting either of them. In contrast, StreamTensor resolves these two conditions through the |itensor|-based typing system, making any dataflow kernels fuseable by design.
\item Due to the reasons above, Stream-HLS did not support the kernel fusion space exploration like StreamTensor, limiting its application on large-scale workloads that cannot be fully deployed on-chip without kernel fusion. For example, Stream-HLS only reports the performance of the multi-head attention layer and feed-forward layer separately, rather than for the entire transformer block.
\end{itemize}

\section{Conclusion and Future Works}

This paper introduces StreamTensor, a compiler framework that automates the generation and optimization of stream-based dataflow accelerators. StreamTensor's main contributions include an |itensor|-based typing system that forms the foundation of the entire framework, a PyTorch-to-device compilation pipeline, and a set of design spaces for exploring key architectural parameters. By addressing common pitfalls in existing frameworks, StreamTensor effectively improves the efficiency of dataflow accelerators. As the demand for efficient AI continues to grow, StreamTensor paves the way for future work in scalable and extensible dataflow compilation.

Looking ahead, StreamTensor's modular design and |itensor| typing system open promising avenues for future work, particularly in extending its compatibility with diverse dataflow architectures and specialized kernel languages. StreamTensor can be adapted to programmable architectures like AMD Versal~\cite{gaide2019xilinx}, Sambanova RDU~\cite{prabhakar2024sambanova}, and Groq LPU~\cite{abts2022groq} by retargeting its low-level compilation and code generation stages. This process would map the dataflow kernels, FIFOs, and layout converters in StreamTensor IR into platform-specific components, such as the AI engines and routing networks in AMD Versal. Similarly, StreamTensor can integrate with kernel languages like Allo~\cite{chen2024allo}, allowing developers to incorporate manually-optimized kernels as black-box components. In both scenarios, the |itensor| system serves as a crucial abstraction layer, enabling StreamTensor to perform high-level dataflow optimizations, including kernel fusion and dataflow component generation, while interfacing with target-specific back-ends and black-box components. This promises to broaden StreamTensor's applicability by leveraging the unique strengths of various hardware platforms and programming languages.

\begin{acks}
We thank all anonymous reviewers, especially our anonymous shepherd, for their valuable feedback and suggestions. We thank Vikram Adve, Jian Huang, and Stephen Neuendorffer for their insightful feedback on this work. We thank Kaiwen Cao for collecting the FPGA experimental results during his internship at Inspirit IoT, Inc. We thank Jinghua Wang for collecting the GPU experimental results. We thank AMD for supporting the FPGA boards to Inspirit IoT, Inc., which are used in this work.
\end{acks}


\bibliographystyle{ACM-Reference-Format}
\bibliography{references}

\end{document}